\documentstyle[12pt,subfigure,graphicx,color,ulem,cancel,cite]{article}
\newcommand{\comment}[1]{}
\newcommand{\newc}{\newcommand}
\newcommand{\disp}{\displaystyle}

\def\issue(#1,#2,#3){{\bf #1}, #2 (#3)}

\def\PREP(#1,#2,#3){Phys.\ Rep. \issue(#1,#2,#3)}
\def\EPJC(#1,#2,#3){Eur.\ Phys.\ J.\ C \issue(#1,#2,#3)}

\def\tanb{\tan\beta}

\def\r2{\sqrt 2}
\def\beq{\begin{equation}}
\def\eeq{\end{equation}}
\def\beqn{\begin{eqnarray}}
\def\eeqn{\end{eqnarray}}
\def\sinW2{\sin^2\theta_W}

\def\mz2{M_{z}^2}
\def\c2b{\cos 2\beta}

\def\m#1{{\tilde m}_#1}

\def\mz{M_Z}
\def\m0{m_0}
\def\mhalf{m_{\frac{1}{2}}}

\def\sec2w{sec^2\theta_W}

\def\gmin2{(g-2)_\mu}

\catcode`\@=11 

\def\lsim{\mathrel{\mathpalette\@versim<}}
\def\gsim{\mathrel{\mathpalette\@versim>}}
\def\@versim#1#2{\vcenter{\offinterlineskip
    \ialign{$\m@th#1\hfil##\hfil$\crcr#2\crcr\sim\crcr } }}

\newc{\wt}{\widetilde}
\newc{\ra}{\rightarrow}
\newc{\s}{\smallskip}
\newc{\nn}{\noindent}
\newc{\non}{\nonumber}

\def \chonep{{\wt\chi_1}^{+}}
\def \chonem{{\wt\chi_1^-}}
\def \chonep2{{\wt\chi_2^+}}
\def \chonem2{{\wt\chi_2^-}}

\def \chonem {{\wt\chi_1^\pm}}
\def \chargino2 {{\wt\chi_2^\pm}}

\def \ch2m {{\wt\chi_2^-}}

\def \chonep {{\wt\chi_1^+}}

\begin{document}
\begin{flushleft}
\end{flushleft}

\begin{center}
{\large \bf Implication of Higgs at 125 GeV within stochastic superspace framework\\}
\vglue 0.5cm
Manimala Chakraborti$^{a}$\footnote{tpmc@iacs.res.in}, 
Utpal Chattopadhyay$^{a}$\footnote{tpuc@iacs.res.in} and 
Rohini M. Godbole$^{b}$\footnote{rohini@cts.iisc.ernet.in}  \\
{$^a$  Department of Theoretical Physics, Indian Association 
for the Cultivation of Science,\\  
2A \& B Raja S.C. Mullick Road, Jadavpur, 
Kolkata 700 032, India}\\
{$^b$  Centre for High Energy Physics, Indian Institute of 
Science, Bangalore 560 012, India
}
\end{center}


\begin{abstract}
We revisit the issue of considering stochasticity of 
Grassmannian coordinates in $N=1$ superspace, which was 
analyzed  previously by Kobakhidze {\it et al}. 
In this stochastic supersymmetry(SUSY) framework, the soft SUSY breaking terms of the minimal
supersymmetric Standard Model(MSSM) 
such as the bilinear Higgs mixing, 
trilinear coupling as well as the gaugino mass parameters are all 
proportional to a single mass parameter $\xi$, a measure of supersymmetry breaking 
arising out of stochasticity. While a nonvanishing trilinear coupling at the high scale 
is a natural outcome of the framework, a favorable signature for obtaining 
the lighter Higgs boson mass $m_h$ at 125 GeV,  
the model produces tachyonic sleptons or 
staus turning to be too light. 
The previous analyses took $\Lambda$, the scale at which input parameters are given,
to be larger than the gauge coupling 
unification scale $M_G$ in order to generate acceptable scalar masses 
radiatively at the electroweak scale. 
Still this was inadequate for obtaining $m_h$ at 125 GeV. 
We find that Higgs at 125 GeV is highly achievable 
provided we are ready to accommodate a nonvanishing scalar mass soft SUSY 
breaking term similar 
to what is done in minimal anomaly mediated SUSY breaking (AMSB) in contrast to a pure AMSB setup.   
Thus, the model can easily accommodate Higgs data, 
LHC limits of squark masses,  WMAP data for dark matter relic density, 
flavor physics constraints and  XENON100 data.
In contrast to the previous analyses we consider 
$\Lambda=M_G$, thus avoiding any ambiguities of a post-grand unified theory physics. The idea of 
stochastic superspace can easily be generalized to various scenarios beyond the MSSM .
\end{abstract}
\noindent
\section{Introduction}
Low energy supersymmetry (SUSY)\cite{SUSYreviews1,SUSYreviews2,SUSYbook,SUSYbook2} 
has been one of the most  promising candidates for a theory of fundamental particles and interactions going beyond the Standard Model (SM); the so-called BSM physics.  The 
minimal extension of the SM including SUSY, namely, the minimal 
supersymmetric Standard Model (MSSM), extends the particle spectrum of the SM by one additional Higgs doublet and the supersymmetric partners of all the SM particles - the 
{\bf s}particles. The Supersymmetric extension of the SM provides a particularly elegant solution to the problem
of stabilizing the electroweak (EW) symmetry breaking scale against large radiative correction and keeps the Higgs ''naturally'' light. In fact,
a very robust upper limit on the mass of the lightest Higgs boson is perhaps one of the important predictions of this theory. Further, this upper limit is linked in an essential way to the values of some of the SUSY breaking parameters in the theory. 
In addition, in $R$--parity conserving SUSY, the lightest supersymmetric particle (LSP) emerges as the natural candidate for the dark matter (DM),the existence of which has been proved beyond any doubt in astrophysical experiments. 
The search for  evidence of the realization of this symmetry in nature(in the context of high energy collider experiments, precision measurements at the high intensity B--factories and in the DM detection experiments) has therefore received enormous attention of particle physicists, perhaps only next to the Higgs boson.  
The recent observation of a boson with mass around $\sim125$ GeV at the Large Hadron Collider (LHC)\cite{HiggsDiscoveryJuly2012} and the rather strong lower limits on the masses of sparticles that possess strong interactions that the LHC searches have yielded\cite{atlas_limits}, 
necessitates careful studies of the  MSSM in the context of all the 
recent low energy data.
In these studies, it is also very  important to seek suitable guiding  principles  
which could  possibly reduce the associated 
large number of SUSY breaking parameters of MSSM. 
Thus, looking for modes of specific SUSY breaking mechanisms that involve only  a few input quantities given at a relevant scale can be useful. Here, 
the  soft SUSY breaking  parameters at the electroweak scale are found via  renormalization group (RG) analyses. Apart from the simplicity of having a few parameters as input, 
such schemes create challenging balancing acts. On the one hand, various 
soft breaking masses and couplings become correlated with 
one another in such schemes. On the other hand, overall one has to accommodate a large number of very  stringent low energy constraints in a comprehensive model 
consisting of only a few parameters. 
A simple and well-motivated example of 
a SUSY model is the minimal supergravity (mSUGRA)\cite{msugra}.  
Here, SUSY is broken spontaneously in a hidden sector and the breaking is communicated 
to the observable sector where MSSM resides via Planck mass suppressed 
supergravity interactions. The model involves soft-SUSY 
breaking parameters like 
(i) universal gaugino mass parameter $\mhalf$,  (ii) the universal scalar 
mass parameter $\m0$ , (iii) the universal trilinear coupling $A_0$,  
(iv) the universal bilinear coupling $B_0$, all given at the gauge 
coupling unification scale. In addition to it, one has the 
superpotential related Higgsino mixing parameter $\mu_0$ with its  
associated sign parameter. The two radiative electroweak symmetry 
breaking (REWSB) conditions may then be used so as to replace $B_0$ and 
$\mu_0$ with the Z-boson mass $\mz$ and $\tanb$, 
the ratio of Higgs vacuum expectation values. Similar to mSUGRA, one has 
other SUSY breaking scenarios like 
the gauge mediated SUSY breaking 
 and models with anomaly mediated SUSY breaking (AMSB) etc\cite{SUSYbook,SUSYreviews2}.  Apart from direct collider physics 
data, one has to satisfy constraints from  
flavor changing neutral current (FCNC) as well as 
flavor conserving phenomena like the anomalous magnetic 
moment of muon, constraints like electric dipole moments 
associated with CP violations or to check whether there is a proper amount of 
dark matter content in an R-parity 
conserving scenarios or proper neutrino masses in R-parity 
violating scenarios\cite{SUSYreviews1,SUSYreviews2,SUSYbook}. 
A single model is yet 
to be found that can adequately explain various stringent experimental 
results and at the same time possesses a sufficient degree of 
predictiveness. Howevr, it is always important to continue the 
quest of a simple model and check the degree of agreement with 
low energy constraints.  

    In this work, we pursue a predictive theory 
of SUSY breaking by considering a field theory 
on a superspace where the Grassmannian coordinates are 
essentially fluctuating/stochastic\cite{Kobakhidze:2008py,Kobakhidze:2010ee,
Kobakhidze:2012kn}. 
We note that in a given SUSY breaking scenario, our limitation of knowing the 
actual mechanism of breaking SUSY is manifested in the soft parameters.
Here, in stochastic superspace framework 
we assume that a manifestation of an unknown but a fundamental mechanism 
of SUSY breaking may effectively lead to 
stochasticity in the Grassmannian parameters of the superspace. With a 
suitably chosen probability distribution, this causes  
a given K\"ahler potential and a superpotential to lead 
to soft breaking terms that 
carry signatures of the stochasticity. As we will see, 
the SUSY breaking is parametrized 
by $\xi$ which is nothing but $1/{<\bar \theta \bar \theta>}$, where the 
symbol $<>$ refers to averaging over the Grassmannian coordinates. The other 
scale that is involved is $\Lambda$. Values of various soft parameters at this scale are the input parameters of the scheme.
The values of the same at the electroweak scale are then obtained from these input values by using the renormalization group evolution.
Considering the superpotential of MSSM, the soft 
terms obtained are readily recognized as the ones supplied by the 
externally added soft SUSY breaking terms of constrained MSSM (CMSSM)\cite{SUSYbook}, except that 
the model is unable to produce a scalar mass soft 
term\cite{Kobakhidze:2008py}. 
Reference \cite{Kobakhidze:2008py} used $\Lambda$ and $\xi$ as free parameters 
while analyzing the low energy signatures within MSSM. $\Lambda$ was chosen  
between $M_G$ to $M_P$, the scale of gauge coupling unification and 
the Planck mass scale respectively.
The model as given in Ref.\cite{Kobakhidze:2008py} is called as 
stochastic supersymmetric model (SSM) and it 
is characterized by universal gaugino mass parameter 
$\mhalf$, universal trilinear soft SUSY 
breaking parameter $A_0$, and  universal bilinear soft SUSY 
breaking parameter $B_0$,
all being related to $\xi$, the parameter related to 
SUSY breaking. We note that with the 
bilinear soft SUSY parameter being given, 
$\tan\beta$, becomes a derived quantity. 

However, as already mentioned,  in spite of the fact that the SSM  
generates soft SUSY breaking terms, it produces no scalar mass 
soft term.   Scalar masses start from zero at  the scale 
$\Lambda$ and renormalization group evolution is used to generate 
scalar masses at the electroweak scale $M_Z$. 
Scalar masses at $M_Z$ severely 
constrain the model because typically scalars are very light. In particular, quite often sleptons turn to be the 
lightest supersymmetric particles or even become tachyonic over a large part of the parameter space.  This is partially ameliorated  when one takes  the high scale $\Lambda$ 
to be larger than the gauge coupling unification scale 
$M_G\sim 2 \times 10^{16}$~GeV. However, in spite of obtaining 
valid parameter space that would provide us with a lightest neutralino  as a possible dark matter candidate in  
R-parity preserving framework, we must note that the low 
values that one obtains for the masses of the first two generation of squarks are hardly something of an  advantage in view of the constraints coming from FCNC  as well as those from the LHC data\cite{atlas_limits}.   
On the other hand, SSM has a natural advantage of being associated with 
a nonvanishing trilinear coupling that is favorable to 
produce a relatively light spectra for a given value of 
Higgs boson mass $m_h$. The recent announcement from the CMS and ATLAS 
Collaborations of the 
LHC experiment about the discovery of a Higgs-like boson 
at $\sim125$~GeV\cite{HiggsDiscoveryJuly2012} thus 
makes this model potentially attractive. However, as explored 
in Ref.\cite{Kobakhidze:2012kn}, SSM as such is unable to 
accommodate such a large $m_h$ in spite of having a built-in feature of 
having a nonvanishing $A_0$. It could at most reach 116 GeV for $m_h$\cite{Kobakhidze:2010ee,Kobakhidze:2012kn} and the constraint due to 
$Br(B_s \rightarrow \mu^+ \mu^-)$ as used in Ref.\cite{Kobakhidze:2012kn} 
was much less stringent in comparison to the same of present 
day\cite{bsmumuexperimental}. Furthermore, it is also important to 
investigate the effect of the direct detection rate of dark matter 
as constrained by the recent XENON100 data\cite{Aprile:2012nq}.

    In this analysis, we would like to give all the input parameters 
at the grand unification scale $M_G$, the scale at which the Standard Model gauge
group, namely, $SU(3) \times SU(2) \times U(1)$, comes into existence. 
Any evolution above $M_G$ would obviously demand choosing 
a suitable gauge group; a question that is not going to be addressed in this work. 
In this way we would like to avoid unknown issues arising 
out of a post-grand unified theory(GUT)\cite{postGut} physics. However, we would rather 
try to meet the phenomenological 
demand of confronting the issue of sleptons becoming tachyonic or avoiding 
scalar masses to become light in general in a {\it minimal modification} 
by considering an externally given scalar mass soft parameter $m_0$ as a 
manifestation of an additional origin of SUSY breaking. 
It would be useful to have the first two generations of scalar masses adequately heavy
so as to overcome the FCNC related constraints and 
LHC data\cite{atlas_limits} on squark masses. Additionally 
this will also be consistent with having the 
lighter Higgs boson mass ($m_h$) to be in the vicinity of 125 GeV. We 
will henceforth denote the model as {\bf Mod-SSM}. 

We may note that traditionally 
minimal versions of models of SUSY breaking have been extended 
for phenomenological reasons. It is also 
true that extending a minimal model often 
lowers predictiveness and may even cause 
partial dilution of the main motivations associated with the building 
of the model. For example, considering 
nonuniversal gaugino or scalar mass scenarios 
may be more suitable than CMSSM or mSUGRA so as to obtain a relatively 
lighter spectra in the context feasibility of exploring via LHC. 
Another example may be given in the context of the minimal AMSB model. 
As we know, a pure AMSB scenario\cite{AMSBorig} is associated with 
form invariance of the renormalization group equations (RGE) of  
scalar masses and absence of flavor violation. However, it produces  
tachyonic sleptons.  In the minimal AMSB model\cite{mAMSB,SUSYbook} one 
introduces an additional common 
mass parameter $m_0$ for all the scalars of the theory. 
This ameliorates the tachyonic slepton problem but it is 
true that we sacrifice the much cherished feature of form invariance 
and accept some degree of flavor violations at the end. We would 
like to explore a non-minimal scenario of 
stochastic supersymmetry model, namely, Mod-SSM  in this spirit. 
We particularly keep in 
mind that the stochastic supersymmetry formalism 
may be used not only within the MSSM framework but it may be extended to 
superpotentials beyond that of the MSSM\cite{Kobakhidze:2010ee}. 
The fact that the 
model with a minimal modification can easily accommodate 
the recent Higgs boson mass range by its generic feature of having a 
nonvanishing trilinear coupling parameter makes it further attractive.  
We believe that our approach of considering an 
additional SUSY breaking scalar mass term is justified  for 
phenomenological reasons. 

   Thus, in Mod-SSM we consider the input of 
soft term parameters $\mhalf$, 
$B_0$ and $A_0$ (all are either proportional to $|\xi|$ or $\xi^*$ ) and 
a universal scalar mass soft parameter $m_0$, 
all being given at a suitable scale which we simply choose as 
the gauge coupling unification scale $M_G$ considering the LEP data 
on gauge couplings could be a hint of the existence of grand unification. 
As in Ref. \cite{Kobakhidze:2008py} we would also 
restrict $\xi$ to be real, either positive or negative. 

We clearly like to emphasize that the SSM 
framework produces no scalar mass soft term at a scale $\Lambda$. 
With $\Lambda$ set to the gauge coupling unification scale $M_G$,  
SSM is plagued with tachyonic sleptons. With 
$\Lambda>M_G$ one can avoid tachyonic scalar but there is no 
scope of obtaining the currently accepted Higgs boson mass. 
In Mod-SSM we consider $\Lambda=M_G$ and add 
the extra scalar mass term for an additional SUSY breaking effect. 
With the trilinear coupling parameter $A_0$ and the bilinear coupling 
parameter $B_0$ becoming correlated with $\mhalf$, Mod-SSM parameter space is essentially 
a subset of the one for the constrained MSSM (CMSSM).  
In Mod-SSM, the SSM inspired nonvanishing $A_0$ parameter is suitable for 
producing an appropriately large loop correction to the 
lighter Higgs boson mass while keeping the overall sparticle spectra 
relatively at a lower range.  CMSSM, on the other hand , 
does not pinpoint with any fundamental physical distinction/motivation 
for such a zone of  
parameter space that is capable of producing the correct Higgs mass 
with similarly smaller sparticle mass scale. 
Even if we consider the stochastic 
superspace model in its original form as only a toy idea, 
we believe that it may be 
worthwhile to explore it with minimal modifications in regard to the 
current Higgs boson mass, as well as other phenomenological constraints.  

\section{Stochastic Grassmannian coordinates and SUSY breaking}
As seen in Ref.\cite{Kobakhidze:2008py} we 
consider an $N=1$ superspace where the Grassmannian coordinates $\theta $
and $\bar{\theta}$ are taken to be stochastic in nature. One starts with identifying  the terms involving superfields in the 
superpotential and the  kinetic energy terms that could be used to 
construct the SUSY invariant Lagrangian density for a given model. 
Each term is then multiplied with a probability distribution 
function ${\cal P}(\theta,\bar 
{\theta})$ and integrated over the Grassmannian coordinates appropriately. 
${\cal P}(\theta,\bar{\theta})$ can be expanded into terms involving 
$\theta$ and $\bar{\theta}$ which obviously has a finite number of terms 
because of the Grassmannian nature of $\theta$ and $\bar{\theta}$. One then 
imposes the normalization condition $\displaystyle \int d^2\theta d^2 
{\bar {\theta}} {\cal P}(\theta,\bar{\theta})=1 $ and vanishing of 
Lorentz nonscalar moments like $<\theta>,<\bar {\theta}>, 
<\theta \bar {\theta}>, <\theta^2 \bar {\theta}>$ and 
$<\theta {\bar {\theta}}^2>$. The stochasticity parameter 
$\xi$ is defined as 
$<\theta \theta>=1/\xi^*$.
Here, $\xi$ is a complex parameter with mass dimension unity. 
As computed in 
Ref.\cite{Kobakhidze:2008py}, and as worked out in this analysis explicitly in the Appendix, 
the above leads to the following Hermitian probability distribution. 
\begin{equation}
{\cal P}(\theta,\bar{\theta}) {|\xi|}^2= {\widetilde {\cal P}}(\theta, \bar \theta)
=1+\xi^*(\theta\theta) 
+\xi({\bar \theta} {\bar \theta}) + |\xi|^2 (\theta\theta) 
({\bar \theta} {\bar \theta}).
\end{equation}
For the simple case of a Wess-Zumino type of scenario\cite{SUSYbook} where the kinetic term 
is obtained from $\Phi^\dagger \Phi$ and the superpotential is  given as  
$\disp{W=\frac{1}{2}m\Phi^2 + \frac{1}{3}h \Phi^3}$, where $\Phi$ is a chiral superfield,  
one finds that the effect of stochasticity as described above leads to the 
following SUSY breaking term: 
\begin{equation}
-L_{soft}=\frac12 {\xi^* m}  \phi^2 + 
\frac23 {\xi^* h} \phi^3 + H.c. . 
\label{Lsofteqn}
\end{equation}

\noindent
Applying the stochasticity  idea to  the superpotential 
of MSSM, along with considering the effect on the gauge kinetic energy  function, the above formalism leads to the following tree level  soft SUSY breaking parameters to be given at the high scale $\Lambda$:\\
a.
universal gaugino mass parameter $\mhalf=\frac{1}{2}|\xi|$,\\ 
b. universal trilinear soft parameter $A_0=2\xi^*$,\\ 
c. universal bilinear Higgs soft parameter $B_0=\xi^*$. \\
Recall that there is no  scalar mass soft SUSY breaking term in SSM.\\
For convenience we take $\xi$ to be a real positive number with an 
additional input sign$(\xi)$.  If we count the universal gaugino mass 
parameter $\mhalf$ as the independent parameter, we have, 
\begin{equation}
A_0=sign(\xi).4\mhalf, B_0=sign(\xi).2\mhalf. 
\label{neweqn}
\end{equation}
As has already been discussed before, we introduce a 
nonvanishing scalar mass parameter $m_0$ and fix $\Lambda$ 
at $M_G$.\footnote{We note that a  vanishing scalar mass parameter at a post-GUT scale, with RG evolution corresponding to  
an appropriate gauge group, would indeed generate  nonvanishing 
scalar mass terms at the unification scale $M_G$\cite{postGut}.}
Thus with the above extension, the input quantities 
for the stochastic SUSY model are:
\begin{center}
$\mhalf$, $m_0$, sign($\mu$) and sign($\xi$).
\end{center}
We note that the model quite naturally is associated with  
nonvanishing trilinear soft breaking terms. 
As we will see, this is quite 
interesting in view of the recent LHC announcement for 
the Higgs mass range centering around 125 GeV\cite{HiggsDiscoveryJuly2012}. 
In this analysis, 
we will discuss only the case of $\xi<0$
because the other sign of $\xi$ does not produce a 
spectra compatible with the dark matter relic density constraint.

The  requirement of the REWSB then results 
in the following relations at the electroweak scale:
\begin{equation}
\mu^2  =  \displaystyle
-\frac{1}{2} M^2_Z +\frac {m_{H_D}^2-m_{H_U}^2 \tan^2\beta} {\tan^2\beta -1}
+ \frac {\Sigma_1 -\Sigma_2 \tan^2\beta} {\tan^2\beta -1}, 
\label{mueqn}
\end{equation}
and,
\begin{equation}
\sin2\beta =  2B\mu/(m_{H_D}^2+m_{H_U}^2+2\mu^2+\Sigma_1+\Sigma_2) \ ,
\label{maeqn}
\end{equation}
where $\Sigma_i$ denote the one-loop 
corrections\cite{effpotLoopCor1,effpotLoopCor2}.
Here, $B$ refers to the value of bilinear Higgs coupling at the electroweak 
scale which has to be consistent with its given value $B_0$ at $M_G$.
$B_0$ is determined via $\mhalf$ apart from a 
sign of the stochasticity parameter as mentioned before. 
Consequently, $\tan\beta$  is a derived quantity in the model. 
$B_0$ at the scale $M_G$ and $B$ at the 
electroweak scale are connected via the following RGE written 
here at the one-loop level:
\begin{equation}
{\frac{dB}{{dt}}}= (3\tilde{\alpha}_2{\tilde{m}_2}
+\frac 35\tilde { \alpha}_1
{\tilde{m}_1}) +(3Y_tA_t +3Y_bA_b+Y_\tau A_\tau) \ ,
\label{Beqn}
\end{equation}
where $t=ln(M_G^2/Q^2)$ with $Q$ being the renormalization scale.  
$\tilde  \alpha_i=\alpha_i/(4\pi)$ for $i=1,2,3$ refer to scaled gauge coupling
constants (with $\alpha_1=\frac53 \alpha_Y$) and $\tilde m_i$ for
$i=1,2,3$ are the running gaugino masses. $Y_i$ are
the squared Yukawa couplings, e.g, $Y_t \equiv y_t^2/(4\pi)^2$ where
$y_t$ is the top Yukawa coupling. 
In this analysis, the value of $\tan\beta$ is determined 
via Eqs.\ref{mueqn}-\ref{Beqn} along with 
$B_0=sign(\xi)2\mhalf$ at the scale $M_G$. 
We use SuSpect\cite{SuspectCode} 
for solving the RGEs and obtaining the spectra.
The code takes $\tan\beta$ as an input quantity. Hence, we implement a 
self-consistent method of solution that starts from a guess value of 
$\tan\beta$ resulting into a $B(M_G)$ that in general 
would not agree with the input of $B_0$. Use of a Newton-Raphson 
root finding scheme ensures a fast convergence toward the correct value of 
$\tan\beta$ when $B(M_G)$ matches with the input of $B_0$. 
Here we stress that we do not encounter any parameter point 
with multiple values of $\tan\beta$ in our analysis.\footnote{See 
Ref.\cite{Drees:1991abTanbetaMulti} for such a general possibility in REWSB 
where $B_0$ is given as an input.}
\section{Results}
\begin{figure}[!htb]
\begin{center}
\includegraphics[scale=0.5]{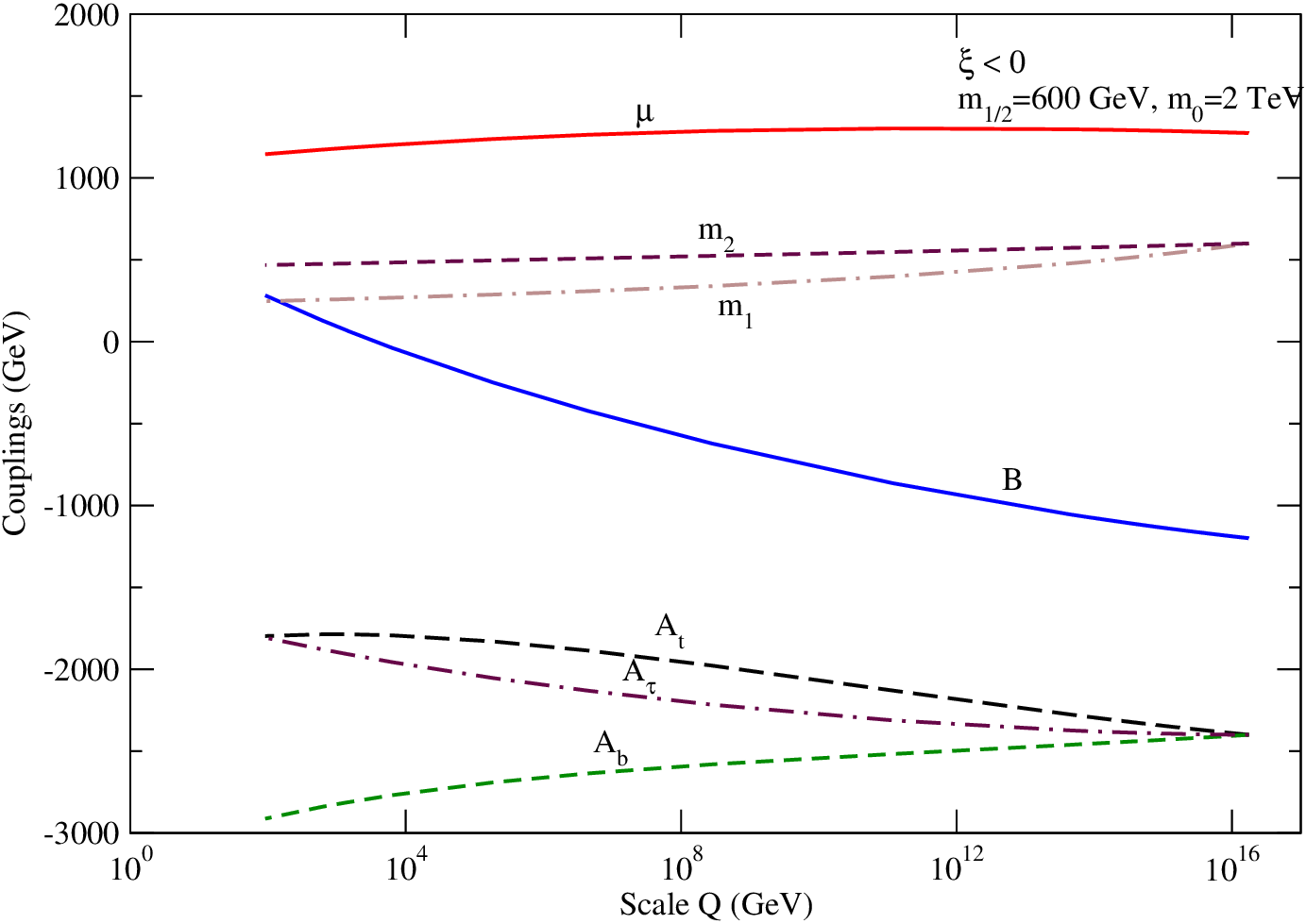}
\caption{{\small
Evolution of a few relevant couplings 
for a specimen input of $m_{1/2}=600$~GeV, $m_0=2$~TeV and $\mu>0$.  
With $\xi<0$, one has $B_0=-2m_{1/2}$ and $A_0=-4m_{1/2}$. 
For a valid parameter point within the model with $\mu>0$ 
we require $B=B(M_Z)>0$, a necessity in order to have a positive 
$\sin2\beta$ from Eq.(\ref{maeqn}).} 
}
\label{rgeneg} 
\end{center}
\end{figure}

The fact that the model has $\tan\beta$ as a derived quantity necessitates 
studying the behavior of the evolution of the bilinear Higgs parameter 
$B$. 
Figure\ref{rgeneg} shows 
the evolution of a few relevant couplings 
for a specimen input of $m_{1/2}=600$~GeV, $m_0=2$~TeV and $\mu>0$ in 
Mod-SSM.  
For $\xi<0$, both $B_0$ as well as $A_0$ are negative, namely, 
$B_0=-2m_{1/2}$ and $A_0=-4m_{1/2}$. 
For a valid parameter point within the model with $\mu>0$ , 
we require $B= B(M_z)>0$, a necessity in order  to have a positive 
$\sin2\beta$ from Eq.(\ref{maeqn}).\footnote{$\tan\beta$ , which is 
the ratio of two vacuum 
expectation values, is positive. Hence, $\sin2\beta={{2\tan\beta} \over 
{(1+\tan^2\beta)}}$ is also positive.}  We note that the denominator in the right-hand 
side of Eq.(\ref{maeqn}) is the square of pseudoscalar Higgs mass 
which needs to be positive.  
 The fact that $B$ is originally 
negative at $M_G$ and has to change to a positive value at $M_Z$ puts 
a strong constraint on the parameter space of the model. Numerically, this 
results in $\tan\beta$ assuming large values. In regard to the evolution of $A$ parameters we defer our discussion until Fig.\ref{mhalf-m0}. \\

\begin{figure}[!htb]
\begin{center}
\includegraphics[scale=0.5]{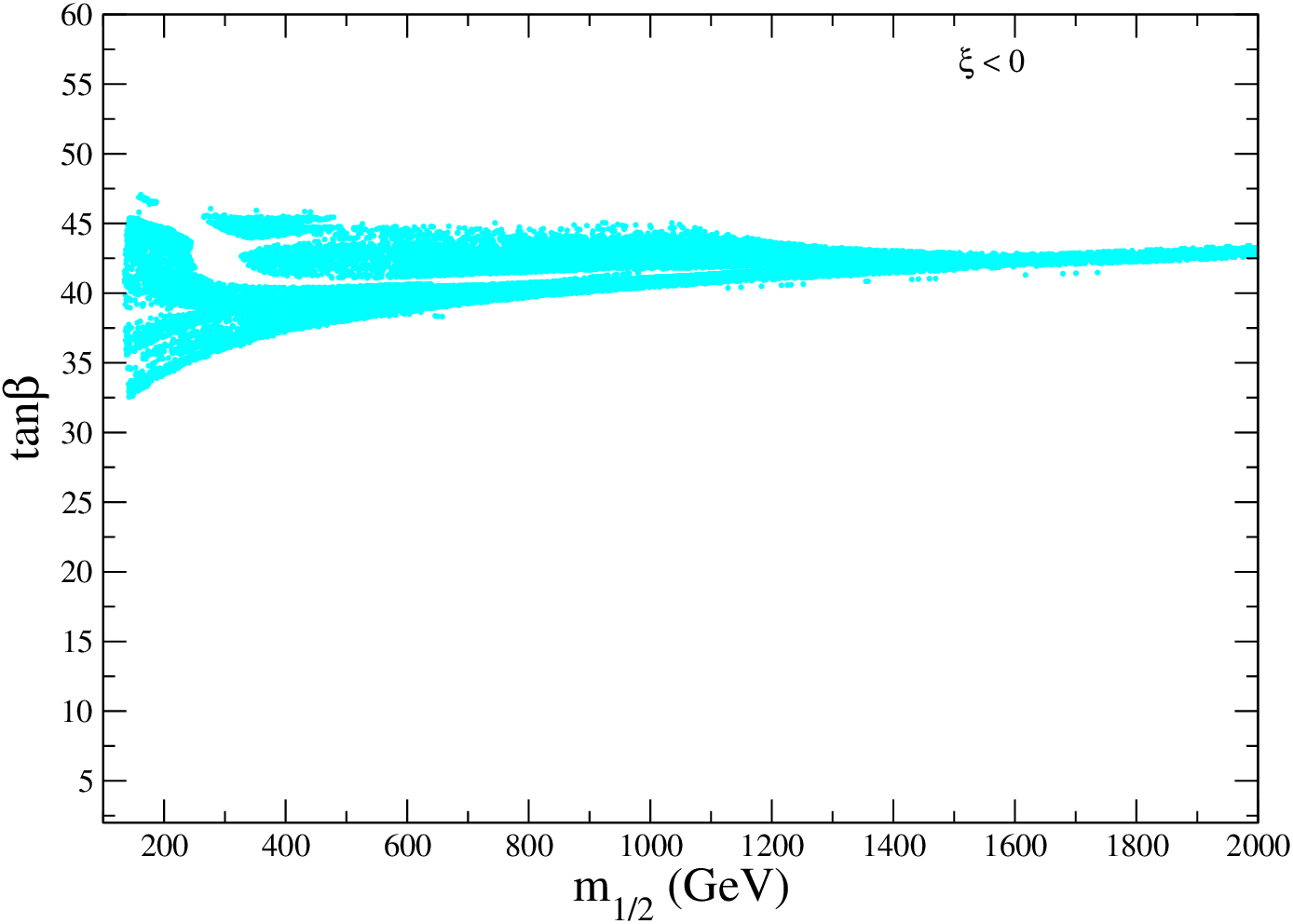}
\caption{{\small
Scatter plot of parameter points in the $\mhalf-\tan\beta$ plane 
when $m_0$ and $\mhalf$ are scanned up to 7 TeV  and 2 TeV, respectively. 
Here we only consider the validity of REWSB constraints of 
Eqs.(\ref{mueqn}) and (\ref{maeqn}). The values 
of $\tan\beta$ that satisfy the REWSB constraint vary from 32 to 48. 
The white region inside the shaded (blue-green) region corresponds to 
invalid parameter points where no consistent solution satisfying 
REWSB could be found even after trying with a large number of iterations. 
The spread of $\tan\beta$ for a given $\mhalf$ arises from 
variation of $m_0$. This spread decreases as $\mhalf$ becomes larger because 
of decreased sensitivity on $m_0$ while satisfying REWSB.}
}
\label{mhalf-tanbeta} 
\end{center}
\end{figure}

\begin{figure}[!htb]
\begin{center}
\includegraphics[scale=0.5]{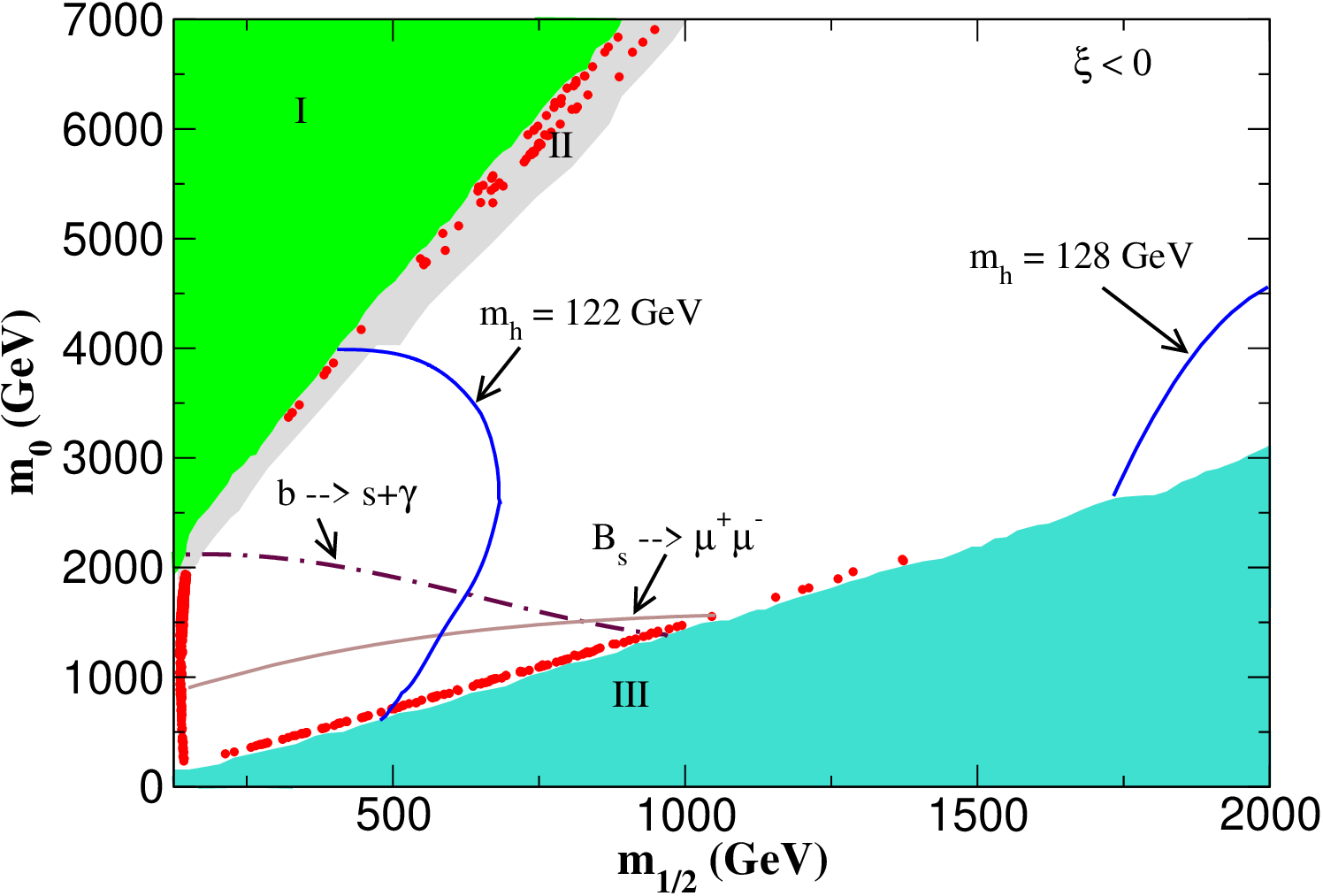}
\caption{{\small
Constraints shown in the $\mhalf-m_0$ plane for $\mu>0$ and $\xi<0$ for 
Mod-SSM. $A_0$ and $B_0$ in the model satisfy 
$A_0=sign(\xi).4\mhalf$ and  
$B_0=sign(\xi).2\mhalf$. $\tan\beta$ becomes a derived quantity that varies 
between 32 and 48. The Higgs boson limits are shown as two solid 
blue lines. $Br(b \rightarrow s  \gamma)$ limit is shown as a
maroon dot-dashed line. The lower part corresponds to discarded 
region via Eq.(\ref{bsgammalimits}) where the branching ratio goes 
below the lower limit of the constraint. 
$Br(B_s \rightarrow \mu^+ \mu^-)$ limit is 
shown as a brown solid line of 
which the lower region exceeds the upper limit of Eq.(\ref{bsmumulimit}). 
The top green region (I) corresponds to discarded zone via REWSB. 
The bottom blue-green region (III) refers to the zone where 
stau becomes LSP or tachyonic. 
The gray region (II) has discontinuous patches of valid parameter 
zones, the details of which are mentioned in the text.  
Red points/areas falling in region-II satisfy the WMAP-7 data 
only for the upper limit of Eq.\ref{wmap7data}. Typically, the red points 
bordering region-III and some part of the extreme left red points (for 
very small $\mhalf$) satisfy both the upper and the lower limits of 
Eq.(\ref{wmap7data}).} 
}
\label{mhalf-m0}
\end{center}
\end{figure}

\noindent
Figure \ref{mhalf-tanbeta} shows a scatter
plot of parameter points in the $\tan\beta-\mhalf$ plane that 
satisfy the REWSB constraints of  Eqs.(\ref{mueqn}) and (\ref{maeqn}). 
As $m_0$ is varied up to 7 TeV and $\mhalf$ up to 2 TeV,  
$\tan\beta$ is seen to have a range of 32 to 48. 
The spread of $\tan\beta$ for a given $\mhalf$ arises from 
variation of $m_0$. 
For smaller values of $\mhalf$, there is a larger dependence of 
$m_0$ on $\tan\beta$. 
Hence, there is a larger spread of 
$\tan\beta$ when $m_0$ is varied. For larger $\mhalf$, the valid 
solution of $\tan\beta$ has a lesser dependence on $m_0$. 
Hence, for larger $\mhalf$ values the spread in values of 
$\tan\beta$ decreases.  
The white region embedded within the blue-green area corresponds to 
parameter points with no valid solution satisfying REWSB. Typically, 
the two REWSB conditions are nonlinear in nature with respect to 
$\mhalf$ and $m_0$. The code unsuccessfully 
tries with a large number of iterations 
to find consistent $\mu^2$ and $m_A^2$ solutions for parameter 
points within the white region (for all $m_0)$. Thus a lack of 
a valid $\tan\beta$ for any value of $m_0$ 
results in the above white region.
It is worth mentioning that the range of valid $\tan\beta$ is much 
larger for the case of $\xi>0$ 
where $B$ stays positive throughout the range from 
$M_G$ to $M_Z$. This is unlike the case of $\xi<0$ under discussion,
where $B$ is 
negative at $M_G$ and necessarily has to become positive 
at the electroweak scale, thus adding stringency to $\tan\beta$ in 
its range. However, as already mentioned,  we will not discuss the case of 
$\xi>0$ 
further because of the resulting overabundance of dark matter for 
the entire parameter space for this  sign of $\xi$. \\
\\
\noindent
We will study now the effect of low energy constraints particularly in the 
context of the recent discovery of the Higgs-like boson\cite{HiggsDiscoveryJuly2012}.
Figure .\ref{mhalf-m0} 
shows the result in the $\mhalf-m_0$ plane for $\xi<0$. 
A sufficiently nonvanishing $A_t$ (see, for example,
Figure \ref{rgeneg}) helps in producing a 
large loop correction to the lighter CP-even Higgs boson $h$. 
This can be understood by looking at the expression for the dominant part of  loop correction to the Higgs boson mass coming from the top-stop sector\cite{Djouadi:2005gj,HiggsOrig1,HiggsOrig2}
\begin{equation}
\Delta m_h^2=\frac{3{\bar{m_t}}^4}{2\pi^2 v^2 \sin^2\beta}
\left[{\rm log}\frac{M^2_S}{{\bar{m_t}}^2} + \frac{X_t^2}{2 M_S^2}
\left(1-\frac{X_t^2}{6 M_S^2}\right) \right ].
\label{higgscor}
\end{equation}
Here,
$M_S=\sqrt{m_{\tilde t_1}  m_{\tilde t_2}}$, 
$X_t=A_t-\mu \cot\beta$, $v=246$~GeV and $\bar {m_t}$ is the  
running top-quark mass that also takes into account 
QCD and electroweak corrections. 
The loop correction is maximized if $X_t={\sqrt 6} M_S$. Clearly, a 
nonvanishing $A_0$ can be useful to increase $\Delta m_h^2$ so that 
$m_h$ reaches the LHC specified zone without a need 
to push up the average sparticle mass scale by a large amount. 
As particularly mentioned in Ref\cite{Kobakhidze:2008py}, 
the SSM is additionally attractive in this 
context since it naturally possesses a nonvanishing 
and large $|A_0|$. We further  note that in this model, the 
lighter Higgs boson $h$ has couplings similar to those in the Standard Model  because the CP-odd Higgs boson mass ($m_{A}$) 
is in the decoupling zone\cite{decouplingHiggs}.  
The ATLAS and CMS results for the possible Higgs boson masses are 
$126.0 \pm 0.4~({\rm stat})\pm 0.4 ({\rm syst})$~GeV and 
$125.3 \pm 0.4~({\rm stat})\pm 0.5 ({\rm syst})$~GeV ,
respectively\cite{HiggsDiscoveryJuly2012}.
In regard to MSSM light Higgs boson mass, we note that there is about 
a 3 GeV uncertainty arising out of uncertainties in the top-quark mass, 
renormalization scheme, as well as scale dependence 
and uncertainties in higher order loop corrections up to three loop\cite{Arbey:2012dq,Heinemeyer:2011aa,Allanach:2004rh,Degrassi:2002fi,higgs3loop}.
Hence, in this analysis we consider the following limits for 
$m_h$:
\begin{equation}
122~{\rm GeV}< m_h <128~{\rm GeV}.
\label{higgslimits}
\end{equation} 
\vspace{0.5cm}
We will outline the other relevant limits used in this analysis. 
A SUSY model parameter space finds a strong constraint from 
$Br(b \rightarrow s  \gamma)$. In SM, the principal contribution 
that almost saturates the experimental 
value comes from the loop comprising of top-quark 
and W-boson\cite{bsgammaSMoriginals}.   
In MSSM, 
principal contributions arise from  
loops containing top quark and charged Higgs bosons, and the same containing 
top squarks and charginos\cite{bsgammaSUSYorigEtc}. 
The chargino loop contributions are proportional to $A_t\mu$ and this 
may cause cancellations or enhancements between the principal 
terms of the MSSM contribution depending on the sign of $A_t\mu$.
Similar to mSUGRA, both the SSM and Mod-SSM with $\xi<0$ also 
typically have $A_t<0$.  With $\mu>0$, this primarily 
means cancellation between the chargino and the charged 
Higgs contributions. This leads to a valid region for 
$Br(b \rightarrow s  \gamma)$ for a larger sparticle mass scale compared to 
the case of $\mu<0$. $Br(b \rightarrow s  \gamma)$ constraint thus 
favors the positive sign of $\mu$ by allowing larger areas of parameter 
space.   
We consider 
the experimental value $Br(b \rightarrow s  \gamma)=(355 \pm 24 \pm 9)\times 10^{-6}$\cite{Asner:2010qj}. This results in the following 
$3 \sigma$ level zone as used in this analysis.  
\begin{equation}
2.78 \times 10^{-4}  <Br(b \rightarrow s  \gamma)<4.32 \times 10^{-4}.
\label{bsgammalimits}
\end{equation}
The above constraint is displayed as a maroon dot-dashed line. The left 
region of this line would be a discarded zone.  
Next, the fact that the stochastic model with $\xi<0$ selects appreciably 
large values for $\tan\beta$ necessitates checking the  
$B_s \rightarrow \mu^+ \mu^-$ limit. This is required because 
$B_s \rightarrow \mu^+ \mu^-$ increases with $\tan\beta$ as $\tan^6\beta$ and 
decreases with increase in $m_A$, the mass of pseudoscalar Higgs boson 
as $m_A^{-4}$\cite{bsmumurefs}. We use the recent experimental 
limit from LHCb\cite{bsmumuexperimental}: 
$Br(B_s \rightarrow \mu^+ \mu^-)_{exp}=
(3.2^{+1.4}_{-1.2}(\mbox{stat.}) ~^{+0.5}_{-0.3} 
(\mbox{syst.})) \times 10^{-9}$. This is contrasted with the SM evaluation 
$Br(B_s \rightarrow \mu^+ \mu^-)_{SM}=(3.23 \pm 0.27)
\times 10^{-9}$~\cite{Buras:2012ru}.  As in Ref.\cite{Roszkowski:2012nq}, 
combining the errors of the LHCb data along with that of the SM result one finds
the following:
\begin{equation}
0.67 \times 10^{-9} <Br(B_s \rightarrow \mu^+ \mu^-)<6.22 \times 10^{-9}.
\label{bsmumulimit}
\end{equation} 
The upper limit of $Br(B_s \rightarrow \mu^+ \mu^-)$ is 
shown as a brown solid line going across Fig.\ref{mhalf-m0}.
 Parameter values in  the region below this curve 
lead to values of $Br (B_s \rightarrow
\mu^+ \mu^-)$ higher than the above limit.

We also compute $Br(B \rightarrow \tau \nu_\tau)$ in this analysis.   
The SUSY contribution to $Br(B \rightarrow \tau \nu_\tau)$ is 
typically effective for large $\tan\beta$ and small charged 
Higgs boson mass scenarios\cite{b2taunuSUSY}. 
The experimental data from {\it{BABAR}}\cite{Lees:2012ju} reads  
$Br(B^+ \rightarrow \tau^+ \nu_\tau)=(1.83^{+0.53}_{-0.49}(\mbox{stat.}) 
\pm 0.24 (\mbox{syst.})) \times 10^{-4}$. The recent result from 
Belle\cite{Adachi:2012mm} for $B^- \to \tau^- \bar{\nu}_\tau$ that used  the
hadronic tagging method is given by 
$Br(B^- \rightarrow \tau^- \bar{\nu}_\tau)=(0.72^{+0.27}_{-0.25}
(\mbox{stat.}) \pm 0.11 (\mbox{syst.})) \times 10^{-4}$. 
The same branching ratio from Belle
extracted by using a semileptonic tagging method is 
$(1.54^{+0.38}_{-0.37}(\mbox{stat.})  ~^{+0.29}_{-0.31} 
(\mbox{syst.})) \times 10^{-4}$\cite{Hara:2010dk}.
We use Ref.\cite{Altmannshofer:2012ks} for the result of averaging of all 
the recent Belle and {\it{BABAR}} data which is $Br{(B\rightarrow \tau \nu)}_{exp}= 
(1.16 \pm 0.22) \times 10^{-4}$. The SM result strongly depends on 
the CKM element $|V_{ub}|$ and the B-meson decay constant. We 
use $Br{(B\rightarrow \tau \nu)}_{SM}= (0.97 \pm 0.22) 
\times 10^{-4}$\cite{Altmannshofer:2012ks}. 
Using the above theoretical and experimental errors appropriately,
we obtain the following:    
\begin{equation}
R_{(B \rightarrow \tau \nu_\tau)}=
\frac{Br{(B \rightarrow \tau \nu_\tau)}_{SUSY}}{Br{(B \rightarrow \tau 
\nu_\tau)}_{SM}}=1.21 \pm 0.30.
\end{equation}
This translates into $0.31 < R_{(B \rightarrow \tau \nu_\tau)}<2.10$ 
at $3\sigma$.
Here $ Br{(B \rightarrow \tau \nu_\tau)}_{SUSY}$ denotes the branching
ratio in a SUSY framework, of course including the SM contribution. 
In general, we find that the model parameter space of Mod-SSM is 
not constrained by $B \rightarrow \tau \nu_\tau$ since charged Higgs 
bosons are sufficiently heavy.

\noindent
We have not, however, included the constraint from muon $g-2$ in this 
analysis considering the tension arising out of large deviation from the
SM value, uncertainty in hadronic contribution evaluations and accommodating 
SUSY models in view of the LHC sparticle mass lower limits\cite{muong-2ETC}.  

\noindent
We now explore the cosmological constraint for neutralino dark matter 
relic density\cite{dmreviews}.  
At 3$\sigma$, the WMAP-7 data\cite{Komatsu:2010fb} are considered as 
shown below:
\begin{equation}
0.094  < \Omega_{{\widetilde \chi}_1^0}h^2<0.128. 
\label{wmap7data}
\end{equation}
The conclusions in regard to the 
relic density constraint is additionally 
found to be sensitive on the top-quark mass in 
this model. We divide the dark matter analysis into two parts depending 
on (a) the top-quark pole mass set at 173.3 GeV and (b) using a spread of 
top-quark pole mass within its range $m_t=173.3 \pm 2.8$~GeV following 
the result of the recent analysis performed in Ref.\cite{Alekhin:2012py}. 
In this context, we note that the experimental value as measured 
by the CDF and D0 Collaborations of Tevatron is:
$m_t^{\rm exp}=173.2\pm 0.9$~GeV\cite{Lancaster:2011wr}.
\footnote{For bottom quark mass we have used ${m_b}^{\overline {\rm MS}}(m_b)=4.19$ GeV.}

\subsection{Analysis with $m_t=173.3$~GeV: Underabundant LSP}   
The lightest neutralino, the LSP of the model 
is typically highly bino dominated except in a few regions where the 
Higgsino mixing parameter $\mu$ turns out to be small. The parameter points 
within all of the white region in Figure \ref{mhalf-m0} have bino-dominated LSP. 
At this point we note that the implementation of the  
REWSB conditions as manifest in Eqs.(\ref{mueqn}) and (\ref{maeqn}) has to be done by
keeping in mind   (i) positivity of  $\sin2\beta$ and  (ii) positivity of $\mu^2$ and
and (iii) the requirement of $B_0, A_0$ being related to  
$\mhalf$ as given by Eq.~(\ref{neweqn}), as well as 
requiring  that the lighter chargino mass lower limit is respected.  
All these requirements lead the top
green shaded region (labeled as I) to be a discarded zone. 
On the other hand, the gray shaded region (II) has discontinuous zones of 
valid parameter points, shown in red. 
The red points have considerably  
small values of $\mu$, thus giving the 
LSP a large degree of Higgsino mixing. 
Apart from the red points, there are no solutions 
in the gray areas of region II.  
The need to satisfy the point (i) as explained above 
along with the requirement to satisfy 
the condition (iii) (implemented via a 
Newton-Raphson method of finding the correct $\tan\beta$) stringently 
negates the existence of  solution zones within region II. As a result, either  
there are solutions with appreciably small $\mu$ or no solution 
at all within this region. 
This on the other hand 
leads to a large amount of $\widetilde \chi_1^0 -\widetilde \chi_1^\pm$ 
coannihilation. 
Such degrees of 
coannihilations indeed cause the LSP to be only a 
subdominant component of DM.
Thus, in this part of the analysis we 
consider the possibility of an underabundant dark matter candidate 
and ignore the lower limit of the WMAP-7 data.
Typically we see that 
the relic density falls below the 
lower limit of Eq.(\ref{wmap7data}) in region II by an order of magnitude. 
Such underabundant LSP scenarios 
have been discussed in several works\cite{underabundantDM}.  

\noindent 
The lower shaded region III is disallowed as the mass of the stau (${\widetilde \tau}_1$) 
turns negative or it is the LSP.
Typically the red strip near region III refers to the 
LSP-stau coannihilation\footnote{See, for example, works in 
Ref.\cite{dmmany} for various annihilation processes in relation to SUSY 
parameter space in general.} zone where the relic density can be 
consistent with both the upper and lower limits of Eq.(\ref{wmap7data}).  
Quite naturally, the coannihilation may be stronger and this would 
additionally produce some underabundant DM points for this zone.  
Finally, the leftmost red region (with very small $\mhalf$)
satisfying WMAP-7 data is discarded by all other constraints. 
We note that only a small region satisfying 
the Higgs mass bound is discarded via $Br(b \rightarrow s \gamma)$. 
On the other hand, the Higgs mass bound line of 122 GeV 
supersedes the constraint imposed by 
the recent data on inclusive search for SUSY by the ATLAS 
experiment\cite{atlas_limits}.  
The most potent constraint to eliminate a large region 
of parameter space with $m_0$ up to 1.5~TeV or so is due to 
$Br(B_s \rightarrow \mu^+ \mu^-)$ data (Eq.(\ref{bsmumulimit})). 
The constraint is effective simply because of the 
large values of $\tan\beta$ involved in the model.
The discarded part of parameter space via the above constraint includes a 
large zone that satisfies the dark matter limit via 
LSP-stau coannihilation.
 
We will now describe the spin-independent 
direct detection scattering cross section for scattering of the LSP with 
proton. The scalar cross section depends on t-channel Higgs 
exchange diagrams and 
s-channel squark diagrams. Unless the squark masses are close to that 
of the LSP, the Higgs exchange diagrams dominate\cite{Drees:1993bu}. 
We note that for the cases of parameter points with 
$\Omega_{\widetilde \chi} h^2 < {(\Omega_{CDM} h^2)}_{\rm min}$, where  
${(\Omega_{CDM} h^2)}_{\rm min}$ refers to the lower limit of 
Eq.(\ref{wmap7data}), 
one must appropriately 
include the fraction of local DM density contributed by the specific 
candidate of DM under discussion while evaluating the event rate.\\

\begin{figure}[!htb]
\begin{center}
\includegraphics[scale=0.5]{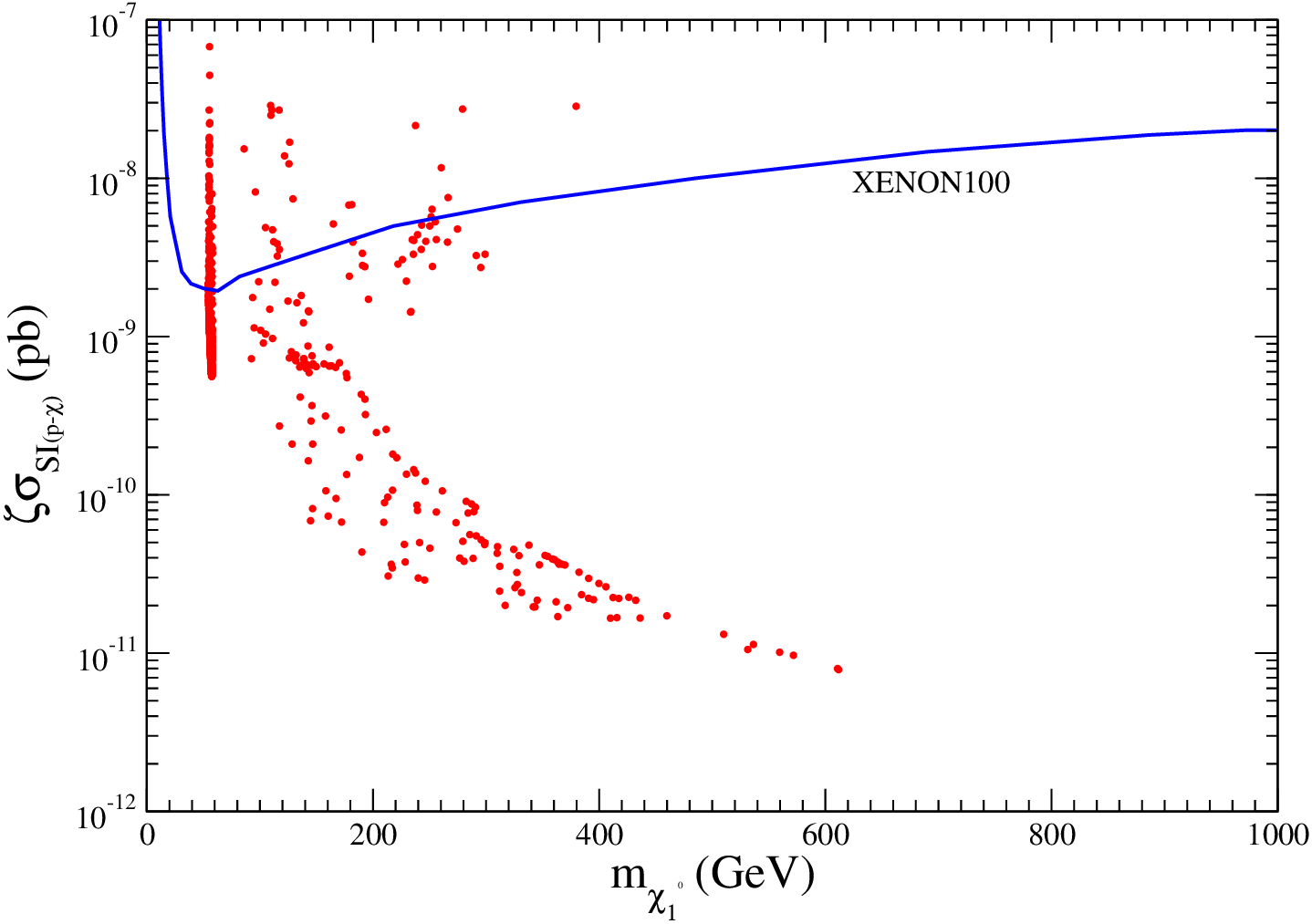}
\caption{{\small Scaled spin-independent ${\tilde \chi}_1^0-p$ 
scattering cross 
section vs LSP mass. The scaling factor is given as $\zeta=\Omega_{\widetilde \chi} h^2/{(\Omega_{CDM} h^2)}_{\rm min}$, where  
${(\Omega_{CDM} h^2)}_{\rm min}$ refers to the lower limit of 
Eq.(\ref{wmap7data}).}}
\label{sigmaSIfigScaled}
\end{center}
\end{figure}
\noindent
This translates into multiplying $\sigma^{SI}_{p{{\widetilde \chi}_1^0}}$ 
for such underabundant scenarios by $\rho_\chi/\rho_0$. Here, $\rho_\chi$ 
is the actual DM density contributed by the specific DM candidate 
contributing to $\rho_0$ where the latter is the local dark matter density. 
We thus use $\rho_\chi=\rho_0 \zeta$  where 
$\zeta=\Omega_{\widetilde \chi} h^2/{(\Omega_{CDM} h^2)}_{\rm min}$. On the 
other hand, $\zeta$ is simply $1$ for abundant or 
overabundant dark matter cases. Thus, we conveniently define 
$\zeta={\rm min}\{1,\Omega_{\widetilde \chi} h^2/{(\Omega_{CDM} h^2)}_{\rm min}\}$\cite{BottinoRescaled}.  
Figure\ref{sigmaSIfigScaled} shows the rescaled cross section as 
computed via micrOMEGAs version 2.4\cite{Belanger:2008sj}. 
We wish to emphasize that while some region of parameter space 
where the LSP typically has a large degree of Higgsino component is 
eliminated via XENON100 data as announced in the summer of 
2012\cite{Aprile:2012nq},  a large section of parameter space 
remains to be explored via future direct detection of DM experiments. 
This consists of both types of 
coannihilation zones, namely, the chargino as well as 
the stau coannihilation zones. 
We must also keep in mind the issue of 
theoretical uncertainties, particularly the hadronic uncertainties 
in evaluating $\sigma^{SI}_{p{{\widetilde \chi}_1^0}}$.
The strangeness content of nucleon finds a large reduction in the evaluation 
of relevant couplings via lattice 
calculations\cite{LatticeStrangenessAndDM}. This is not incorporated 
in our computation while using micrOMEGAs to calculate the cross section.  
Thus, the above itself will cause a reduction of 
$\sigma^{SI}_{p{{\widetilde \chi}_1^0}}$
by almost an order of magnitude. There is also an appreciable amount of 
uncertainty of the local dark matter density\cite{furtherref1}. All these points
need to be kept in mind while evaluating the implications of Fig. 4 for SUSY models.\\

\subsection{Analysis with $m_t=173.3 \pm 2.8$~GeV: LSP of right abundance}
It is to be noted  that in the gray area (shown as region II) of Fig.\ref{mhalf-m0} 
the existence of a valid solution depends very critically on the parameters of the model. 
It was found that either (i) we obtain a very small $\mu$ (barely 
satisfying the lighter chargino mass lower limit) in the gray area 
that would 
only provide us with extreme coannihilation between lighter chargino and 
LSP leading to underabundance of DM, or (ii) we find 
no valid solution at all. The sensitivity arises from the stringency of 
satisfying REWSB on the parameter space for this region. 
In other words, for a given $\mhalf$, a small change in assumed $m_0$ 
for a valid parameter point would 
produce a small change in $\tan\beta$ in the {\it white} region producing 
most probably another valid parameter point.  
However, the change may not be allowed via 
REWSB, particularly Eq.(\ref{maeqn}) if the parameter point is considered 
in the {\it gray} region. Eq.(\ref{maeqn}) means $\sin2\beta$ needs to 
be a positive quantity less than unity.  This typically becomes 
a severe constraint even if condition of satisfying the lighter 
chargino mass lower limit is met.    
With the top-quark mass having a strong influence on REWSB ,
it may be useful to investigate 
whether varying $m_t$ may extract newer valid points within the gray 
region that would have a suitable $\mu$ so as to satisfy 
a well-tempered\cite{Arkani-Hamed:2006mb} LSP situation. This 
will then have the right abundance of dark matter satisfying both 
the lower and the upper limit of DM of Eq.(\ref{wmap7data}).

The present experimental data on top-quark mass read:
$m_t^{\rm exp}=173.2\pm 0.9$~GeV\cite{Lancaster:2011wr}. 
Recently Ref.\cite{Alekhin:2012py} predicted the pole mass of top-quark 
to be $m_t^{\rm pole}=173.3 \pm 2.8$~GeV. The analysis used the 
next-to-next-to-leading order
(NNLO) in the QCD prediction of the inclusive $pp \rightarrow t \bar t +X$ cross 
section and the Tevatron and LHC data of the 
the same cross section. The comparison between the experimental and 
 theoretical results helped extracting the top-quark mass in 
the modified minimal subtraction ($\overline{\rm MS}$) scheme. This was then 
used to compute the pole mass $m_t^{\rm pole}$.  
We now extend our analysis by investigating the effect of varying 
top-quark pole mass within the above range 
($m_t^{\rm pole}=173.3 \pm 2.8$~GeV)
on the solution space of the model. Indeed, we will see that even with a 
variation of $0.9$~GeV, the range of 
experimental error would be enough to have a substantial effect on the conclusions.\\ 
\begin{figure}[!htb]
\begin{center}
\includegraphics[scale=0.5]{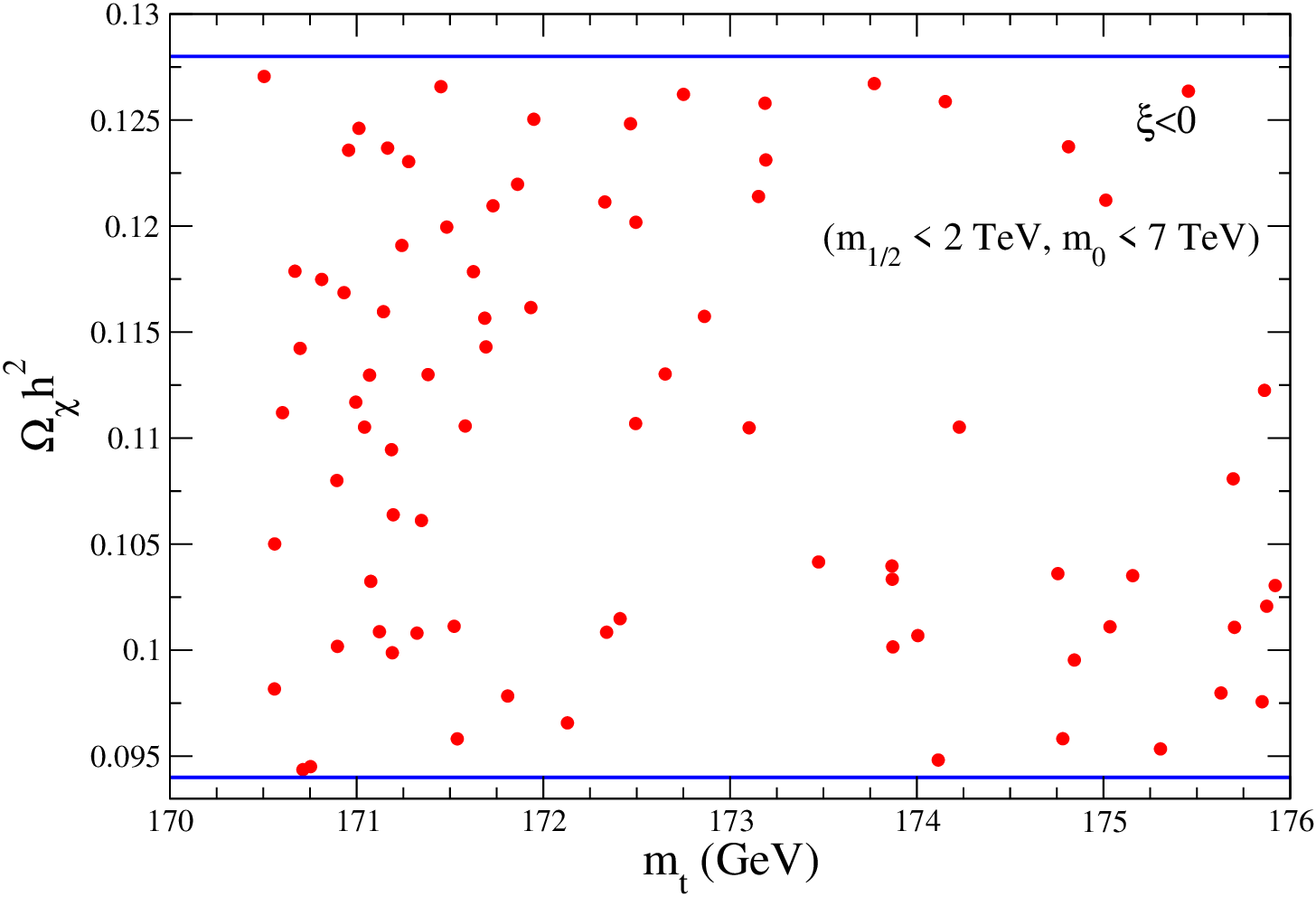}
\caption{{\small Relic density satisfied points shown in red 
that fall within the lower and upper limits of WMAP-7 data for the 
neutralino relic density, when $m_t$ 
is varied by 2.8 GeV on either side of 173.3 GeV. Here 
$\mhalf$ and $m_0$ are scanned up to 2 TeV and 7 TeV, respectively, for 
$\mu>0$ and $\xi<0$. The two blue lines are the WMAP-7 limits of 
Eq.(\ref{wmap7data}). The value of $m_t=173.3$~GeV as used in 
Figure \ref{mhalf-m0} visibly falls in a disfavored zone in the context 
of obtaining the correct relic density, particularly toward the 
lower limit of of Eq.(\ref{wmap7data}).} 
}
\label{top-dependence}
\end{center}
\end{figure}

\noindent
Figure \ref{top-dependence} shows the scattered points 
that satisfy the WMAP-7 given range of relic density 
for the aforesaid variation of $m_t\equiv m_t^{\rm pole}$. 
Here, $\mhalf$ and $m_0$ are varied up to 2 TeV and 7 TeV, respectively. 
We see from this figure that the central value of 
$m_t$~(=173.3)~GeV as 
considered in Figure \ref{mhalf-m0} is indeed the one which has the {\it least 
amount of possibility to satisfy the WMAP-7 data. The occurrence 
of points which satisfies WMAP-7 data  is particularly rare around this value of $m_t$,  
near the lower part of the limit of relic density.} We already found that the gray 
region (region II) of Figure \ref{mhalf-m0} 
is a sensitive zone because of REWSB, where $\mu$ can be quite small.  
For such small values of $\mu$, one can only expect a large degree 
of $\widetilde \chi_1^0 -\widetilde \chi_1^\pm$ 
coannihilation which results into very small 
relic density. The latter goes below the lower 
limit of Eq.(\ref{wmap7data}). Hence no red point exists near the 
bottom blue line of Figure \ref{top-dependence} for this value 
of $m_t$.
On the contrary, a value of $m_t$ less than 
1 GeV from the central value, which is only within the 
experimental error of $m_t^{\rm exp}$, would cause  
to have an LSP with correct abundance for DM. 
The most favored zone for $m_t$, however, would 
be from 171 to 172 GeV for satisfying the relic density 
limits. In fact, a reduced degree of sensitivity for 
satisfying REWSB is the reason for obtaining a well-tempered LSP while 
considering a top-quark mass little away from the central value.
\\
\begin{figure}[!htb]
\begin{center}
\includegraphics[scale=0.5]{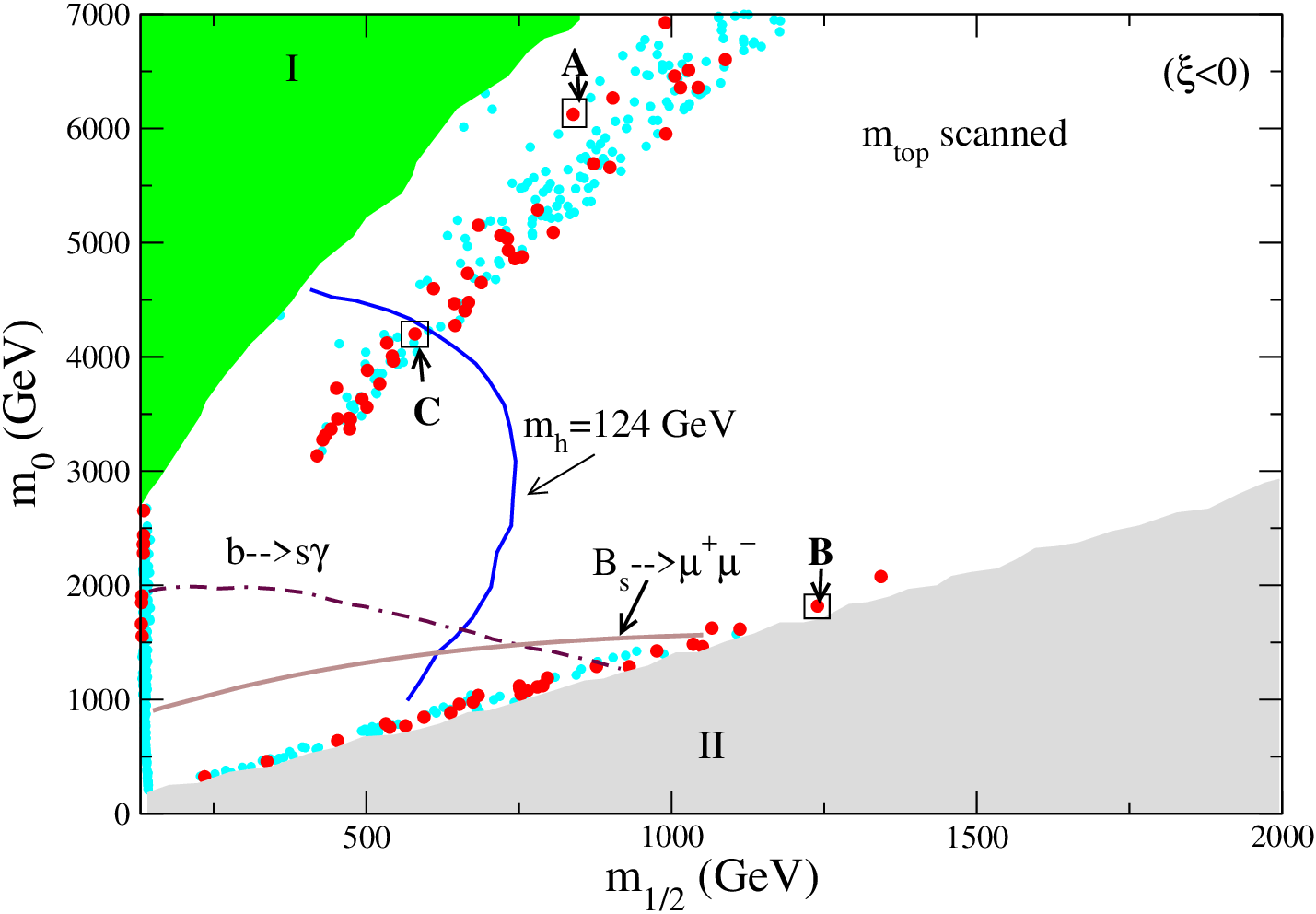}
\caption{{\small Scattered points in the
$\mhalf-m_0$ plane for $\mu>0$ and $\xi<0$ 
when top-quark mass is varied by 2.8 GeV on either side of 173.3 GeV. 
Region I is the same as that in Fig.\ref{mhalf-m0} except that it is 
now a discarded zone for all values of $m_t$ within its limit. 
The region II is discarded via stau turning the LSP or 
turning itself tachyonic for all $m_t$.
The constraints from  
$Br(b \rightarrow s  \gamma)$ 
and $Br(B_s \rightarrow \mu^+ \mu^-)$ are as shown. 
The lines show the boundary of 
purely discarded zones irrespective of variation 
of $m_t$ within its range.  
There is no region toward the left of the blue line labeled as 
$m_h=124$~GeV for which $m_h$ may become larger than 124 GeV irrespective 
of values of $m_t$. Three benchmark points A,B and C are shown 
corresponding to Table~\ref{spectratable}.}
}
\label{m0mhalfwithtopscanned}
\end{center}
\end{figure}

\noindent
Figure \ref{m0mhalfwithtopscanned} shows the effect of scanning the top-quark 
mass on the $\mhalf-m_0$ plane. 
Region I shown in green is a disallowed area of parameter space 
via REWSB similar to Fig.\ref{mhalf-m0}, except that it is 
now an invalid region for all values of $m_t$ within its limit. 
Thus, this region is  smaller in extension  than the corresponding region of 
Figure \ref{mhalf-m0}.  
The region II is disallowed because of stau turning tachyonic or the least massive. 
The constraints from  
$Br(b \rightarrow s  \gamma)$ and $Br(B_s \rightarrow \mu^+ \mu^-)$ are as shown. 
The lines denote the boundary of purely 
discarded zones irrespective of variation 
of $m_t$ within its range.  
The blue line for $m_h=124$~GeV means that $m_h< 124$~GeV for all the region 
left of the line irrespective of the value of $m_t$ within the range.  
The ATLAS specified 
limit\cite{atlas_limits} for squarks also falls well within this left zone. 
The neutralino relic density 
satisfied areas are below region I and above region II. 
The {\it red} points satisfy both the 
limits of the WMAP-7 data, and thus correspond to having 
the right degree of abundance of 
DM. On the other hand, we have also shown {\it blue-green} points that 
only satisfy 
the upper limit of the WMAP-7 data. 
In this part of the analysis, we consider  the
LSP to have the correct abundance so as to be a unique candidate for DM. 

\begin{figure}[!htb]
\begin{center}
\includegraphics[scale=0.5]{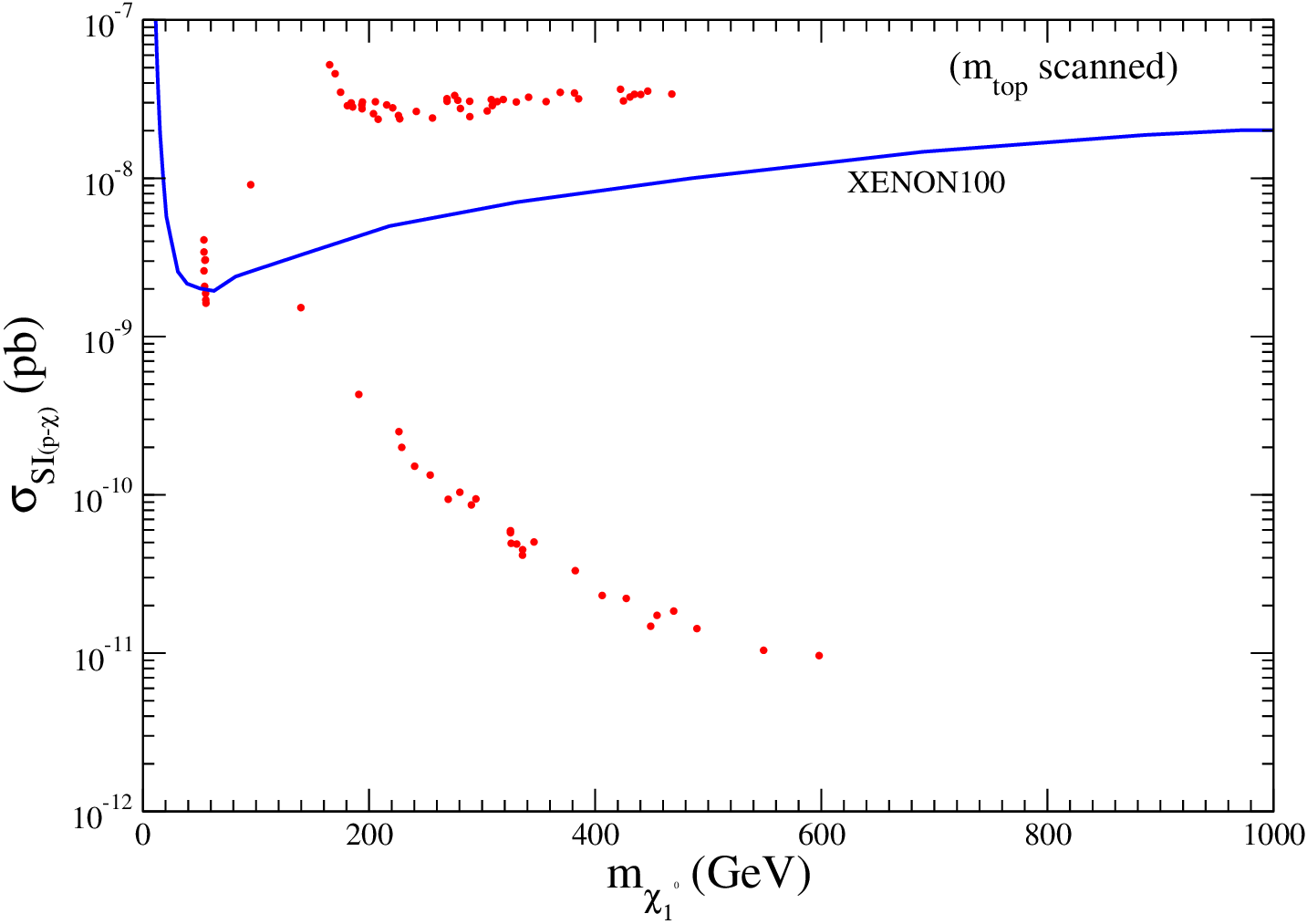}
\caption{{\small Spin-independent LSP-proton scattering 
cross section vs LSP mass 
when $m_t$ is scanned along with $\mhalf$ and $m_0$ for $\mu>0$ and $\xi<0$. 
Only WMAP-7 satisfied points (for both the lower and the upper limit) 
are shown along with the XENON100 exclusion limit.}} 
\label{sigmaSIfigtopscanned}
\end{center}
\end{figure}

\noindent
Finally we show the effect of varying $m_t$ on the 
spin-independent LSP-proton scattering cross section 
in Figure \ref{sigmaSIfigtopscanned}. Only the WMAP-7 satisfied points are 
shown (in red). 
Considering an order of magnitude of 
uncertainty (reduction), primarily because of the issue of 
strangeness content of nucleon as well as astrophysical uncertainties 
as mentioned before, 
we believe that the recent XENON100 data still can accommodate 
Mod-SSM even while considering the LSP as a unique candidate of 
DM.\\

\noindent
Table~\ref{spectratable} shows three benchmark points of the model. The top-quark
 mass $m_t$ is as shown for each of the cases. 
Points A and B correspond to values of 
$m_t$ which are entirely within the experimental error. The points A and 
C correspond to the upper region of Figure \ref{m0mhalfwithtopscanned}, 
and these two points correspond to the hyperbolic branch (HB)/focus point (FP) 
zone\cite{hyperHB,focusFP}. These 
points are associated with a large degree of 
$\widetilde \chi_1^0 -\widetilde \chi_1^\pm$ coannihilation.
The degree of agreement between the desired value of $B_0$ and the one 
obtained via the Newton-Raphson iteration may be seen from the fifth and the 
sixth rows. We allowed a maximum deviation of 1$\%$ in the iterative 
procedure while scanning the parameter space. A majority of the points are 
consistent within 0.1$\%$ in this regard.   
The charginos and the neutralinos are relatively lighter for points A and C in 
comparison to those in point B. The scalar particles, on the other hand ,
are relatively lighter for point-B.  Undoubtedly, like many SUSY models 
analyzed or reanalyzed after the Higgs boson discovery,  
the spectra is on the heavier side. The fact 
that $A_0$ is nonvanishing and adequately large helps in reducing 
the average sparticle mass to a great extent compared to vanishing 
$A_0$ scenarios satisfying the current lighter Higgs boson limit.  The
$m_h$ for point-C, however, goes below the assumed limit 
of Eq.(\ref{higgslimits}). However , we,
still believe that it is within an acceptable zone considering the 
various uncertainties to compute $m_h$ as mentioned before. 
Points A and C have larger spin-independent scattering 
cross section $\sigma_{p\chi}^{SI}$ 
than the XENON100 limit. But, we believe this is within 
acceptable limit considering the existing uncertainties 
arising out of strangeness content of nucleon as well as 
those from astrophysical origins, particularly from local DM density. Finally, 
it is also possible to satisfy all the limits in full subject to a $5\%-10\%$ 
heavier spectra and/or considering a multicomponent 
DM scenario.\\   
\begin{table}[ht]
\begin{center}
\begin{tabular}[ht]{|l|c|c|c|}
\hline
Parameter & A & B & C \\
\hline
\hline
$m_t$ & 	      173.10 &     173.87   &	     171.58 \\
$m_{1/2}$ & 	     838.78 &     1239.16   &	     579.69  \\
$m_0$ &  	    6123.75 &     1817.69   &	    4200.55  \\
$(A_0=-4 m_{1/2})$ & 	   -3355.13 &    -4956.64   &	   -2318.75 \\ 
$(B_0=-2 m_{1/2})$ & 	   -1677.56 &    -2478.32   &	   -1159.37  \\
$B_0 ~(as ~output)$ &  	   -1683.56 &    -2478.32   &	   -1160.27 \\
$\tan\beta~(as ~output)$ &  	      45.86 &       40.92   &	      45.11 \\ 
$sgn(\mu)$ & 	       1 &        1   &	       1 \\
$\mu$ &  	     403.86 &     2508.85   &	     310.43 \\
$m_{\tilde g}$ &  	    2145.53 &     2727.64  &	    1525.80 \\
$m_{\tilde u_L}$ & 	    6247.84 &     2994.87  &	    4292.70 \\
$m_{\tilde t_1},m_{\tilde t_2}$ & 	    3758.76,    4376.60 &     1333.10,    2078.60   &	    2587.56,    3026.20 \\
$m_{\tilde b_1},m_{\tilde b_2}$ & 	    4397.10,    4886.58 &     2054.58,    2339.25  &	    3037.22,    3381.67 \\
$m_{\tilde e_L}, m_{\tilde {\nu_e}}$& 	    6119.53,    6119.05 &     1983.72,    1982.21   &	    4197.61,    4196.89 \\
$m_{{\tilde \tau}_1},m_{\tilde {\nu_\tau}}$ &	    4750.43,    5482.91 &      549.62,    1536.41  &	    3281.31,    3770.25 \\
$m_{{\tilde \chi_1}^{\pm}},m_{{\tilde \chi_2}^{\pm}}$ & 	     406.25,     741.03 &     1038.00,    2491.42  &	     304.84,     518.70 \\
$m_{{\tilde \chi_1}^0},m_{{\tilde \chi_2}^0}$ & 	     356.62,     417.02 &      548.95,    1038.00   &	     241.80,     316.70 \\
$m_{{\tilde \chi_3}^0},m_{{\tilde \chi_4}^0}$ & 	     424.40,     741.06 &     2489.56,    2491.14  &	     322.35,     518.80 \\
$m_A,m_{H^{\pm}}$ &  	    2573.69,    2573.69 &     1967.83,    1967.50   &	    1846.04,    1846.04 \\
$m_h$ &  	     124.42 &      126.55  &	     {\color{red} $\mathit 121.57$} \\
$\Omega_{\tilde \chi_1}h^2$ & 	  0.1105 &   0.1002  &	  0.1106 \\
$BF(b\to s\gamma)$ & 	  $3.23\times {10}^{-4}$ &  $2.96  \times {10}^{-4} $   &	  $3.15\times {10}^{-4}$ \\
$BF(B_s\to \mu^+\mu^-)$ & $2.98 \times {10}^{-9} $ &   
$5.23 \times {10}^{-9}$   &	  $2.89\times {10}^{-9}$ \\
$R_{(B\to \tau \nu)}$ & 	  0.98 &   0.98  &	  0.97 \\
$\Delta a_\mu$ & 	  $5.78 \times {10}^{-11}$ &   $1.65 \times {10}^{-10}$  &	  $1.22\times {10}^{-10}$ \\
$\sigma_{p\chi}^{SI}$ & 	  {\color{red} $\mathit 3.05 \times {10}^{-8}$} &   $1.04 \times {10}^{-11}$   &	  {\color{red} $\mathit 2.64\times {10}^{-8}$} \\
\hline 
\end{tabular}
\end{center}
\caption{Spectra of three specimen parameter points A, B and C as shown in 
Fig.\ref{m0mhalfwithtopscanned}. Results with marginal deviation from the 
assumed limits are shown in italics (red). 
Muon $g-2$ is not imposed as a constraint in this analysis. 
$B_0=B(M_G)$ has two entries. The first one is the desired value while  
the second one is the value obtained with a suitable $\tan\beta$ as  
found from the Newton-Raphson root finding scheme. See text for 
further details.  
}
\label{spectratable}
\end{table}
\clearpage

\section{Conclusion}
In the SSM\cite{Kobakhidze:2008py}, 
one assumes that a manifestation of an unknown but a fundamental mechanism 
of SUSY breaking may effectively lead to 
stochasticity in the Grassmannian parameters of the superspace.
With a suitable probability distribution decided out of 
physical requirements, stochasticity in Grassmannian coordinates for  
a given K\"ahler potential and a superpotential may lead 
to well-known soft breaking terms. When applied to the superpotential of 
the MSSM, the model leads to soft breaking terms like 
the bilinear Higgs coupling term, a trilinear soft term as well as a 
gaugino mass 
term, all related to a parameter $\xi$, the scale of SUSY breaking. The other 
scale considered in the model of Ref.\cite{Kobakhidze:2008py} 
at which the input quantities are given 
is $\Lambda$, where the latter can assume a value between the gauge 
coupling unification scale $M_G$ and the Planck mass scale $M_P$. 
There is an absence of 
a scalar mass soft term at $\Lambda$ in the original model that leads to stau 
turning to be the LSP or even turning itself tachyonic if $\Lambda$ is 
chosen as $M_G$. This is only partially ameliorated when $\Lambda$ 
is above $M_G$. However, the model because 
of its nonvanishing trilinear parameter is 
potentially accommodative to have a larger lighter Higgs boson mass
via large stop scalar mixing. As recently shown in 
Ref.\cite{Kobakhidze:2012kn} the model in spite of its nice feature 
of bringing out the desired soft breaking terms of MSSM is not able to produce 
$m_h$ above 116 GeV, and it has an LSP which is only a subdominant component 
of DM.  

    In this work, a minimal modification (referred as Mod-SSM) 
is made by allowing a nonvanishing 
single scalar mass parameter $m_0$ as an explicit soft breaking term. 
Phenomenologically the above addition is similar to what was considered 
in minimal AMSB while confronting the issue of sleptons turning 
tachyonic in a pure AMSB framework.  The modified model 
successfully accommodates the lighter Higgs boson mass near 125 GeV. 
Additionally,
it can accommodate the stringent constraints from the dark matter 
relic density, $Br(B_s \rightarrow \mu^+ \mu^-)$, $Br(b\to s\gamma)$ 
and XENON100 data on direct detection of dark matter. A variation of 
top-quark mass within its allowed range is included in the analysis and 
this shows the LSP to be a suitable candidate for dark matter satisfying
both the limits of WMAP-7 data. Finally, we remind that the idea of 
stochastic superspace can easily be generalized to various scenarios beyond the MSSM .

{\bf 
\noindent
Acknowledgments}\\
M.C. would like to thank the Council of Scientific and
Industrial Research, Government of India for support.
U.C. and R.M.G are thankful to the CERN THPH division (where the work was initiated) for its hospitality .
  R.M.G.  wishes to acknowledge the Department of 
Science and Technology of India, for financial support under Grant no.
SR/S2/JCB-64/2007.  We  would like to thank B.~Mukhopadhyaya, S. Kraml, 
P.~Majumdar, K.~Ray, S.~Roy and S.~SenGupta for valuable discussions.

\section{Appendix}
Using Ref.\cite{Kobakhidze:2008py}, we 
consider the following Hermitian probability distribution :
\begin{eqnarray}\nonumber
{\cal P}(\theta,\bar{\theta})= A+\theta^\alpha\Psi_\alpha 
+{\bar \theta}_{\dot \alpha}{\bar \Xi}^{\dot \alpha} +\theta^\alpha\theta_\alpha B+{\bar \theta}_{\dot \alpha}
{\bar\theta}^{\dot \alpha} C+\theta^\alpha{\sigma^\mu}_{\alpha{\dot \beta}}{\bar \theta}^{\dot \beta}
V_\mu\\
+\theta^\alpha\theta_\alpha{\bar \theta}_{\dot \alpha}{\bar \Lambda}^{\dot \alpha} 
+{\bar \theta}_{\dot \alpha}{\bar\theta}^{\dot \alpha}\theta^\alpha\Sigma_\alpha
+\theta^\alpha\theta_\alpha{\bar \theta}_{\dot \alpha}{\bar\theta}^{\dot \alpha} D
\label{origProbability}
\end{eqnarray}
Here, A, B, C, D and $V_\mu$ are complex numbers. 
$\Psi$, ${\bar \Xi}$, ${\bar \Lambda}$, and
$\Sigma$ are Grassmann numbers.  

\noindent
In order to arrive at the results of Ref.\cite{Kobakhidze:2008py}, we use the 
following\cite{SUSYbook}: $d^2\theta= -\frac{1}{4}d\theta^\alpha 
d\theta_\alpha$, $d^2 {\bar \theta}= -\frac{1}{4}d{\bar \theta}_{\dot \alpha} 
d{\bar \theta}^{\dot \alpha}$, $d^4\theta=d^2\theta d^2 {\bar \theta}$, $ \int d^2\theta ~(\theta \theta) = 1$,
$\int d^2\bar \theta ~(\bar \theta \bar \theta) = 1 $, $\int d^2\theta = \int d^2\bar \theta = 0 $ and $\int d^2\theta ~{\theta^\alpha} = \int d^2 \bar \theta ~{\bar \theta}_{\dot \alpha} = 0 $. 

\noindent
{\bf Normalization:}\\
First, 
${\cal P}(\theta,\bar{\theta})$ should 
satisfy the normalization condition 
$\int d^2\theta d^2 {\bar {\theta}} ~{\cal P}(\theta,\bar{\theta})=1$. 
All the terms except the one with the 
coefficient $D$ vanishes in $\displaystyle \int d^2\theta d^2 {\bar {\theta}} ~{\cal P}(\theta,\bar{\theta})$. Thus $D=1$. 

\noindent
{\bf Vanishing fermionic moments:}\\
Next, we require vanishing of moments of fermionic type because of the 
requirement of Lorentz invariance. The fact that 
$<\theta^\beta>=\displaystyle \int d^2\theta d^2 {\bar {\theta}} 
~\theta^\beta {\cal P}(\theta,\bar{\theta})=0$ means $\Sigma_\alpha=0$. 
Similarly, $<\bar {\theta_{\dot \beta}}>=0$ means ${\bar \Lambda}^{\dot \alpha}=0$,  and $<\theta^\beta \bar {\theta_{\dot \gamma}}>=0$ leads to $V_\mu=0$. 
Finally, $<\theta^2 \bar {\theta_{\dot \beta}}>=0$ gives ${\bar \Xi}^{
\dot \alpha}=0$ and $<\theta^\beta {\bar \theta}^2>=0$ gives $\Psi_\alpha=0$.

\noindent
{\bf Bosonic moments:}\\
\noindent 
We now compute the bosonic moments. \\
$<\theta \theta> = \displaystyle \int d^2\theta d^2 {\bar \theta} 
~\theta^\beta \theta_\beta {\cal P}(\theta,\bar \theta)=\displaystyle \int d^2\theta d^2 {\bar \theta} ~(\theta \theta)(\bar \theta \bar \theta) C=C$.  \\
Similarly, $<\bar \theta \bar \theta>=B$. Calling $B=1/\xi$, one has 
$B=C^*=1/\xi$. 
\\Furthermore,$<\theta \theta \bar \theta \bar \theta>=A$. The 
fact that $<\theta \theta \bar \theta \bar \theta>=<\theta \theta><\bar \theta \bar \theta>$ leads to $A=1/{|\xi|}^2$.

\noindent
Thus we find the following Hermitian probability measure for the stochastic Grassmann variables:
\begin{equation}
{\cal P}(\theta,\bar{\theta}) {|\xi|}^2={\widetilde {\cal P}}(\theta,\bar{\theta})
=1+\xi^*(\theta\theta) 
+\xi({\bar \theta} {\bar \theta}) + |\xi|^2 (\theta\theta) 
({\bar \theta} {\bar \theta}).
\label{actualProbability}
\end{equation}

\noindent
We consider the Wess-Zumino model with a single chiral superfield $\Phi$. $\Phi$ 
has the following expansion\cite{SUSYbook} : 
\begin{eqnarray}\nonumber
\Phi &=& \phi(x) 
-i\theta\sigma^\mu\bar\theta\partial_\mu\phi(x)
- \frac{1}{4}\theta^2{\bar\theta}^2\partial_\mu\partial^\mu\phi(x)
+\sqrt{2}\theta\psi(x) \\
& &
+\frac{i}{\sqrt{2}}\theta^2\partial_\mu\psi(x)\sigma^\mu\bar \theta 
+\theta^2 F(x).  
\end{eqnarray}
Correspondingly, for $\Phi^\dagger$ we have:
\begin{eqnarray}\nonumber
\Phi^\dagger &=& \phi^*(x) 
+i\theta\sigma^\mu\bar\theta\partial_\mu\phi^*(x)
- \frac{1}{4}\theta^2{\bar\theta}^2\partial_\mu\partial^\mu\phi^*(x)
+\sqrt{2}{\bar \theta}{\bar \psi(x)} \\
& &
-\frac{i}{\sqrt{2}}{\bar \theta}^2 \theta \sigma^\mu \partial_\mu{{\bar 
\psi}(x)} 
+{\bar \theta}^2 F^*(x).  
\end{eqnarray}
The kinetic term of the Lagrangian ${\cal L }$ 
is ${\left[\Phi^\dagger \Phi\right]}_D$. One finds,
\begin{eqnarray}
\Phi^\dagger \Phi &=&{|\phi|}^2 + \sqrt 2 \theta\psi \phi^* 
+\sqrt 2 \bar \theta \bar \psi \phi
 + \theta^2 \phi^*F
+ {\bar \theta}^2 F^*\phi+2\bar \theta \bar \psi \theta \psi  \nonumber \\
& &
+i\sqrt2 \theta^2 \bar \theta {\bar \sigma}^\mu \psi{[\partial_\mu]}\phi^*  
+\sqrt 2 \theta^2 \bar \theta \bar \psi F
- 2i\theta \sigma^\mu \bar \theta \phi^*{[\partial_\mu]}\phi \nonumber \\
& &
+i\sqrt 2 {\bar \theta}^2 \theta \sigma^\mu\bar \psi {[\partial_\mu]}\phi
+\sqrt 2 {\bar \theta}^2 \theta \psi F^*  \nonumber \\
& &
+\theta^2
{\bar \theta}^2 \left(F^* F +\frac12 \partial_\mu \phi^* {[\partial^\mu]}\phi 
-\frac12 \phi^* {[\partial_\mu]} \partial^\mu \phi 
+i\psi\sigma^\mu{[\partial_\mu]}\bar \psi\right).
\label{kineticInfull}
\end{eqnarray}
Here, $X{[\partial^\mu]}Y=\frac12\left(X\partial_\mu Y -Y \partial_\mu X \right)$. Upon vanishing the  appropriate surface terms, the 
D term in particular reads:
\begin{equation}
{\left[\Phi^\dagger \Phi\right]}_D=F^*F + \partial_\mu \phi^* \partial^\mu 
\phi+\frac{i}{2}\left(\psi \sigma^\mu \partial_\mu \bar \psi- \partial_\mu 
\psi \sigma^\mu \bar \psi\right).
\label{actualkinetic}
\end{equation} 

\noindent
Next, we consider the superpotential $W$ given by  
$W=\frac12 m \Phi^2+ \frac13 
h \Phi^3$.  One has:
\begin{eqnarray}
\Phi^2 &=& \phi^2 + 2\sqrt 2\theta \psi \phi +
\theta^2(2 \phi F-\psi\psi), \nonumber ~{\rm and}\\ 
\Phi^3 &=& \phi^3+3\sqrt 2 \theta \psi \phi^2 
+3\theta^2 (F\phi^2-\psi \psi \phi). 
\end{eqnarray}
Thus,
\begin{equation}
W=\left(\frac12 m\phi^2 +\frac13 h \phi^3\right) +\sqrt 2 \theta\psi (m\phi++h\phi^2)
+\theta^2 \left(m \phi F -\frac12 m \psi \psi + h F \phi^2 -h \psi \psi \phi
\right).
\label{superpotentialexpr}
\end{equation}
The potential energy density term will be as follows: 
\begin{equation}
{[W+ H.c.]}_F= \left(m \phi F -\frac12 m \psi \psi + h F \phi^2 
-h \psi \psi \phi\right)+H.c.
\label{potentialdensity}
\end{equation}
{\bf Kinetic and potential terms averaged over $\theta$ and $\bar \theta$ and emergence of soft SUSY breaking terms:}\\
Averaging over the Grassmannian coordinates, we compute 
$L=<{\cal L}> 
=\displaystyle \int d^2 \theta d^2 \bar \theta 
{\widetilde {\cal P}}(\theta,\bar \theta) {\cal L}$. 
Here ${\cal L}$ is the usual 
super-Lagrangian density: ${\cal L}=\Phi^\dagger \Phi + W
\delta^{(2)}(\bar \theta) + W^\dagger \delta^{(2)} (\theta)$.  
Then, using Eq.(\ref{kineticInfull}) and Eq.(\ref{actualProbability}) 
$<{\cal L}_{\rm kinetic}>$, namely, the kintetic part of $L$ is 
found as,
\begin{equation}
<{\cal L}_{\rm kinetic}>= {\left[\Phi^\dagger \Phi\right]}_D + 
\overbrace{{|\xi|}^2{|\phi|}^2 + \xi^* \phi F^* + \xi \phi^* F}^{\cancel{SUSY}}.
\label{derivedKE}
\end{equation} 
Similarly, the potential energy density averaged over $\theta$ and $\bar \theta$ 
is given by,
\begin{equation}
<W + H.c >=\displaystyle \int d^2 \theta d^2 \bar \theta 
{\widetilde {\cal P}}(\theta,\bar \theta) \left(W
\delta^{(2)}(\bar \theta) + W^\dagger \delta^{(2)} (\theta)\right).
\end{equation}
Using Eq.(\ref{superpotentialexpr}) and Eq.(\ref{actualProbability}) we find,
\begin{equation}
<W + H.c.>=\left[
\overbrace{\xi^*\left(\frac12 m\phi^2 +\frac13 h \phi^3\right)}^{\cancel{SUSY}}+
\left(m \phi F -\frac12 m \psi \psi + h F \phi^2 
-h \psi \psi \phi\right)
\right]+H.c.
\label{derivedPE}
\end{equation}
Total Lagrangian is then: 
\begin{equation}
L= <{\cal L}>=<{\cal L}_{\rm kinetic}> + <W + H.c.>.
\label{totlagrangiansum}  
\end{equation}

\noindent
Using Eqs.(\ref{potentialdensity}), (\ref{derivedKE}) and (\ref{derivedPE})
we break $L$ into SUSY invariant and SUSY breaking parts as follows:
\begin{equation}
 L={<{\cal L}>}_{\rm SUSY} + 
{<{\cal L}>}_{\rm \cancel{SUSY}},
\end{equation}
where
\begin{equation}
{<{\cal L}>}_{\rm SUSY} ={\left[\Phi^\dagger \Phi\right]}_D +
{[W+ h.c.]}_F, 
\end{equation}
and 
\begin{equation}
{<{\cal L}>}_{\rm \cancel{SUSY}} ={|\xi|}^2{|\phi|}^2 + \xi^* \phi F^* + \xi \phi^* F +\left[\xi^*\left(\frac12 m\phi^2 +\frac13 h \phi^3\right)+h.c. \right]. 
\end{equation}
The equations of motion of auxiliary fields are then
\begin{eqnarray}
F &=& -(\xi^* \phi +m\phi^* +h {\phi^*}^2) \nonumber, ~{\rm and} \\
F^* &=& -(\xi \phi^* +m\phi +h {\phi}^2) .
\label{auxiliaryEqns}
\end{eqnarray}
Substituting $F$ and $F^*$ in $L$, one finds,
\begin{equation}
L=L_{\rm On-shell-SUSY}+ L_{\rm soft},
\end{equation}
where $L_{\rm On-shell-SUSY}$ is the usual on shell SUSY invariant Lagrangian
for interacting Wess-Zumino model and is given by,
\begin{eqnarray}
L_{\rm On-shell-SUSY}&=& \partial_\mu \phi^* \partial^\mu 
\phi+\frac{i}{2}\left(\psi \sigma^\mu \partial_\mu \bar \psi- \partial_\mu 
\psi \sigma^\mu \bar \psi\right)-m^2{|\phi|}^2 -h^2{({|\phi|}^2)}^2 \nonumber \\
& &
-\left[  
\left(
mh{|\phi|}^2\phi+\frac12 m\psi\psi +h \psi\psi \phi
\right)+H.c.
\right],
\end{eqnarray}
and $L_{\rm soft}$ is given by, 
\begin{equation}
-L_{\rm soft}=\left[
\left(
\frac12 \xi^*m\phi^2 + \frac23 h \xi^* \phi^3
\right)+H.c.
\right].
\label{lsofteqn}
\end{equation}
We remind that a negative sign in the left-hand side of 
Eq.\ref{lsofteqn} appears simply because of considering a positive 
sign before $<W + H.c.>$ in Eq.(\ref{totlagrangiansum}) while writing 
the total Lagrangian. 
We note that it is only the superpotential term in the original theory 
that leads to soft breaking terms in the resulting Lagrangian 
($m \rightarrow \xi^* m$ and $h \rightarrow 2 \xi^* h$ going from $W$ to $L_{\rm soft}$). 
Thep presence of a vector field will not lead to any soft SUSY breaking term.
This can easily be seen by considering a vector superfield in Wess-Zumino 
gauge. \\

\noindent
{\bf MSSM:}\\
In the MSSM, as mentioned in Ref.\cite{Kobakhidze:2008py},
the superpotential term $\mu H_u H_d$ will lead to 
$\xi^* \mu {\tilde H}_u {\tilde H}_d$ and terms like 
${\hat y}^{up}Q U^c H_u$ will lead to $2\xi^* {\hat y}^{up} \tilde Q 
{\tilde U}_c {\tilde H}_u$ as soft SUSY breaking terms. 
Here, fields with tildes denote the scalar 
component of the corresponding chiral superfields. One further obtains 
a gaugino mass term $\frac{\xi^*}{2}\Sigma_i \lambda^{(i)} \lambda^{(i)} $.
Thus, one finds a universal gaugino mass $m_{1/2}=\frac12|\xi|$, 
a bilinear Higgs soft parameter $B_\mu=\xi^*$ and a 
universal trilinear soft parameter  
$A_0=2\xi^*$. With no resulting scalar mass term, one has the universal 
scalar mass parameter $m_0=0$. These are the input quantities to be given at 
a scale $\Lambda$. Low energy spectra are then found via RG evolutions. 
Reference \cite{Kobakhidze:2008py} considered $\xi$ and $\Lambda$ as the 
input quantities and considered $M_G< \Lambda <M_P$.

\end{document}